# Systematic Fluorination is a Powerful Design Strategy Towards Fluid Molecular Ferroelectrics

Calum J. Gibb[1*], Jordan Hobbs [2], Richard. J. Mandle [1,2]

[1]School of Chemistry, University of Leeds, Leeds, UK, LS2 9JT

[2]School of Physics and Astronomy, University of Leeds, Leeds, UK, LS2 9JT

*Author for correspondence: c.j.gibb@leeds.ac.uk

**Abstract**

Ferroelectric nematic ($N_F$) liquid crystals combine liquid-like fluidity and orientational order of conventional nematics with macroscopic electric polarization comparable in magnitude to solid state ferroelectric materials. Here, we present a systematic study of twenty-seven homologous materials with various fluorination patterns, giving new insight into the molecular origins of spontaneous polar ordering in fluid ferroelectric nematics. Beyond our initial expectations, we find the highest stability of the $N_F$ phase to be in materials with specific fluorination patterns rather than the maximal fluorination which might be expected based on simple models. We find a delicate balance between polar and apolar nematics which is entirely dictated by the substitution of the fluorine atoms. Aided by electronic structure calculations, we show this to have its origins in the radial distribution of charge across the molecular surface, with molecules possessing a more oscillatory distribution of electrons across their surfaces possessing a higher propensity to form polar nematic phases. This work provides a new set of ground rules and designing principles which can inform the synthesis of future ferroelectric nematogens.

# Introduction

Through its applications in display devices, the conventional nematic (N) phase (**Figure 1a**) underpinned a revolution in display technology since the mid 1980's. The ferroelectric nematic ($N_F$) phase was recently discovered in 2017 [1,2] and combines the orientational order of conventional nematic liquid crystals with polar ordering, resulting in a 3D fluid with bulk electric polarization whose magnitude is comparable to solid-state ferroelectric materials (**Figure 1b**) [3,4]. The discovery of the $N_F$ phase at equilibrium has garnered significant scientific interest due it's the potential to 'remake science and technology' [5–10]. The $N_F$ phase combines fluidity with a large spontaneous polarisation value resulting in non-linear optical properties [11,12] and significant electric field screening potential [13]. Together, these point to a plethora of possible end-uses including electrooptic devices [14–16], production of entangled photon pairs [17], tuneable lasers [18] and reflectors [19] to name but a few possible applications.

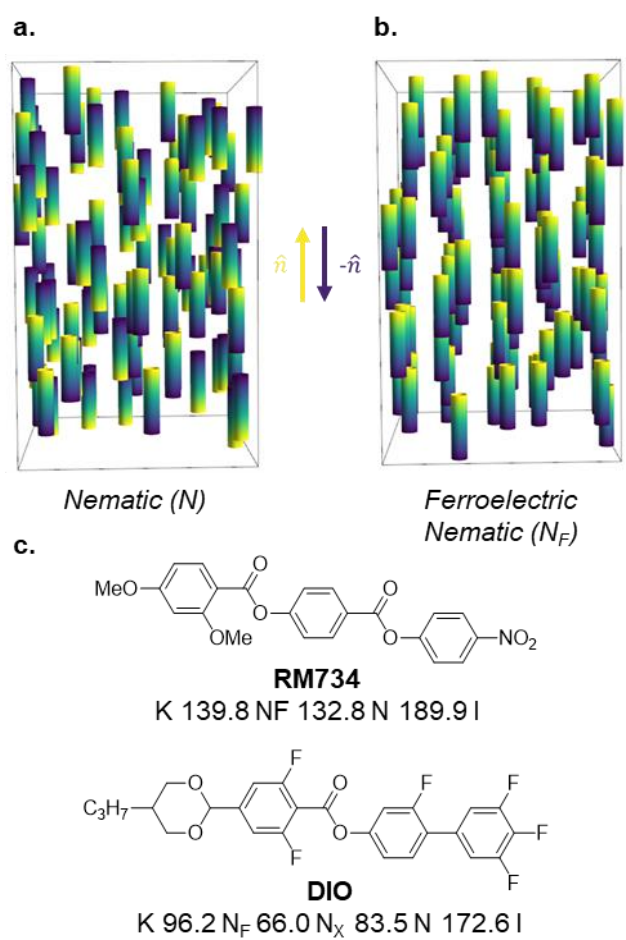

**Figure 1.**   Schematic representations of **(a)** the apolar nematic (N) phase and **(b)** the ferroelectric nematic ($N_F$) phase. Both phases only have orientational ordering with molecules aligning along a unit vector termed the director ($\hat{n}$). In the polar, $N_F$ case, the molecular electric dipole moments of the molecules spontaneously align, resulting in a phase possessing a macroscopic polarisation (i.e. $-\hat{n} \neq \hat{n}$) – this is not observed in the conventional, apolar N phase where molecules can freely

rotate about the short molecular axis (i.e $\hat{n} = \hat{n}$) ; and **(c)** the chemical structures of the archetypal ferroelectric nematic materials, **RM734** [1] and **DIO** [2], with their associated transition temperatures (°C).

While the rich physics of the $N_F$ phase is rightly celebrated, the molecular basis of this new state of matter is often overlooked. Archetypal materials, such as **RM734** [1] **DIO** [2] (**Figure 1c**), have typically been the subjects of most physical investigations but are non-ideal for practical applications due to their propensity to suffer from irreversible structural changes at moderate temperature [20–22]. The scope of studies into the structure-property relationship within the context of the $N_F$ phase to date have been narrow, largely focusing on changes to molecular length, terminal chain length, and small changes in fluorination of the two archetypal materials [4,23,24] . We considered that by presenting an exhaustive study into fluorination patterns in a simple biphenyl benzoate liquid crystal, we could generate a new structure space that shows the $N_F$ phase while also probing the delicate balance between polar and apolar ordering.

The chemical structure-property relationships governing the molecular origins of the $N_F$ phase are still relatively unknown. To date, most molecules which exhibit the $N_F$ phase all possess significant molecular electric dipole moments (μ) (circa. 8 D), although there is still debate about the role dipole moments play in the formation of the $N_F$ phase [4,12,25–27]. To this end, we elected to design a new chemical structure space such that the position of all fluorine atoms are additive to the overall longitudinal molecular electric dipole moment, systematically increasing the number and position of the substituents in-order to screen all possible fluorination patterns of our chosen structure type (**Scheme 1**). This culminated in the systematic synthesis of twenty-seven homologues which possess moderate to large values of μ (**1-27**). Full synthetic details, including spectroscopic and purity data, can be found in the ESI to this article. For simplicity, we refer to compounds **1-27** by the acronym ***X·Y·Z*** where X, Y and Z refer to the number of fluorine substituents on each aromatic ring, beginning with the nitrile bearing ring and ending with the benzoate (for example, the most fluorinated materials synthesised (**1**), is given the acronym **2·2·2**).

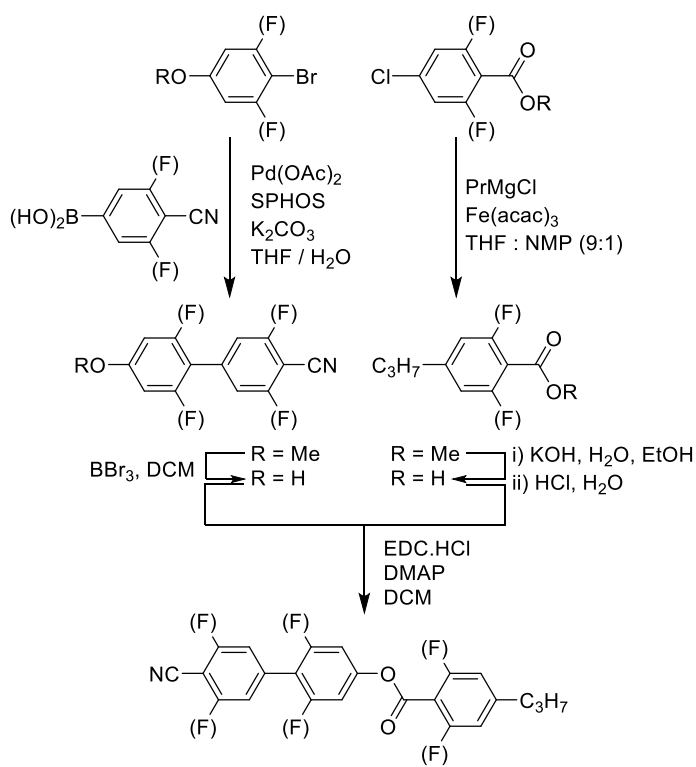

**Scheme 1.** The synthetic route used in the synthesis of materials **1-27**.

## Results and Discussion

Materials **10-27** contain the least F atoms of the materials studied (F < 4) and exhibit solely conventional nematic behaviour with the exception of **2** which also exhibits a SmA phase (**Table S1**). A simple inspection of the nematic to isotropic (I) phase transition temperature ($T_{N-I}$) reveals the expected trend whereby increasing the number of F atoms generally leads to a decrease in the values of $T_{N-I}$. This simply reflects the changes in free volume afforded by additional F atoms inhibiting the efficient packing of the molecules into the N phase. Gratifyingly, increasing the number of fluorine substituents yields materials with more interesting mesomorphic behaviour (**1-9**, **Figure 2** and **Figure 3a**). **2.2.2 (1)** displays a monotropic $N_F$ phase at 133.5 °C, which forms directly from the isotropic liquid. The $N_F$ phase was identified firstly by polarized optical microscopy (POM) by the appearance of a characteristic banded texture (for example see **Figure 3b(i)**) followed by the conformation of the transition temperature by differential scanning calorimetry (DSC) (**Figure S1**). The polar nature of the $N_F$ phase was confirmed by a single peak in the current response (**Figure 3c**). Specifically for the direct I-$N_F$ phase transition, in the isotropic phase a pre-transition field induced I-$N_F$ phase transition as seen from the double peaks in the current trace due to the critical-like first order nature of the I-$N_F$ transition (**Figure S2**) [28]. **2.2.2** also shows an immediate saturation of the spontaneous polarisation, indicating a strongly first order transition from complete isotropy to a $N_F$ phase (**Figure 3d**). X-ray scattering measurements confirmed the assignment of the $N_F$ phase where diffuse signals are seen in both the wide and small angle regions indicating orientational ordering of the molecules with no positional order, respectively, across the entire phase range (for example see **Figure 3e**).

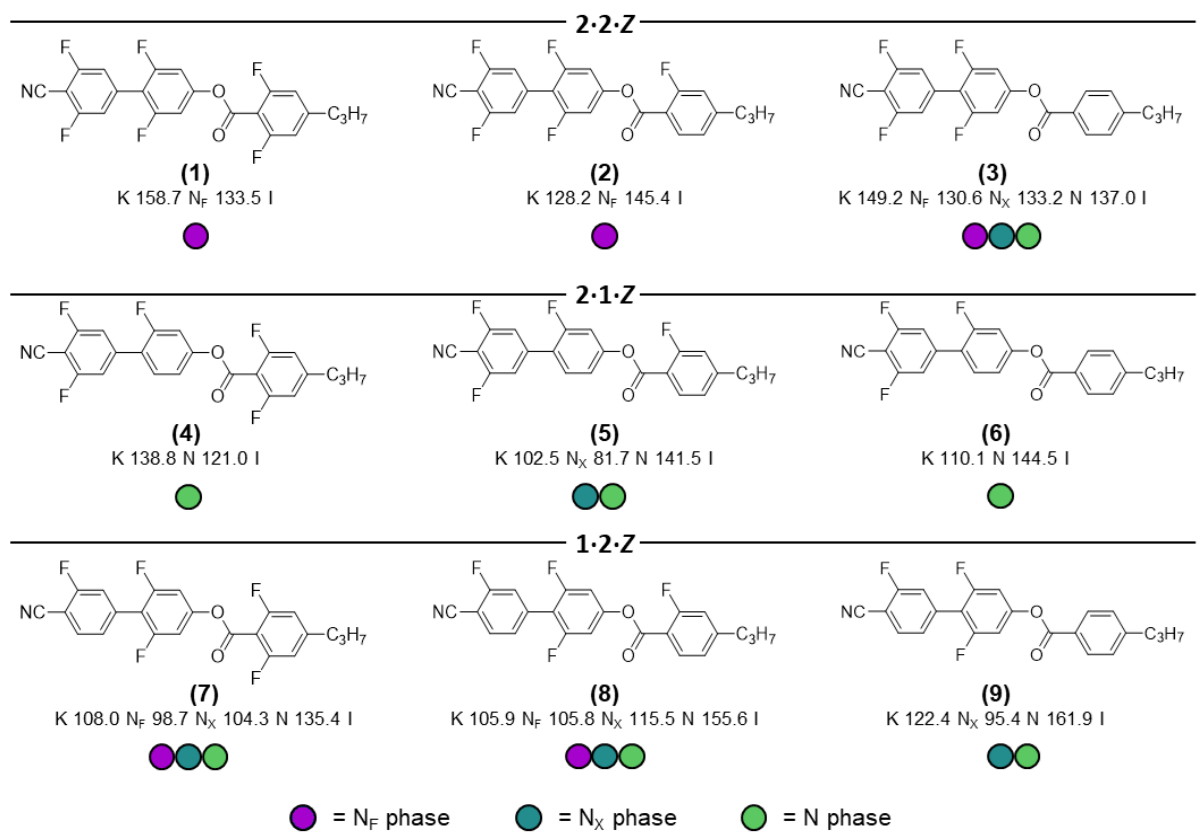

**Figure 2.** The chemical structures, phase sequences and associated transition temperatures (°C) of materials **1-9**. The analogous data for **10-27** may be found in the ESI. K = melting point; $N_F$ = ferroelectric nematic; $N_X$ = antiferroelectric nematic; N = nematic; I = isotropic liquid.

Surprisingly, removal of a single fluorine atom (to afford **2.2.1** (**2**)) leads to a significant increase in the $N_F$-I transition temperature ($T_{NF-I}$) resulting in the $N_F$ phase observed for **2.2.1** being enantiotropic, despite this modification leading to a decrease molecular electric dipole moment (µ) (**Figure 3f**). Interestingly, this appears to be a general trend when comparing **1-9** whereby, regardless of the fluorination pattern and the phase sequence of the material, homologues with *X·Y·1* fluorination patterns have significantly higher transition temperatures associated with polar order (i.e $N_F$-$N_X$ or $N_X$-N) than their more fluorinated counterparts whilst possessing smaller values of µ. Considering the current understanding of the molecular origins of the $N_F$ phase, one might assume maximal fluorination would result in the most desirable materials. This therefore makes this result rather unexpected and something that we will revisit shortly.

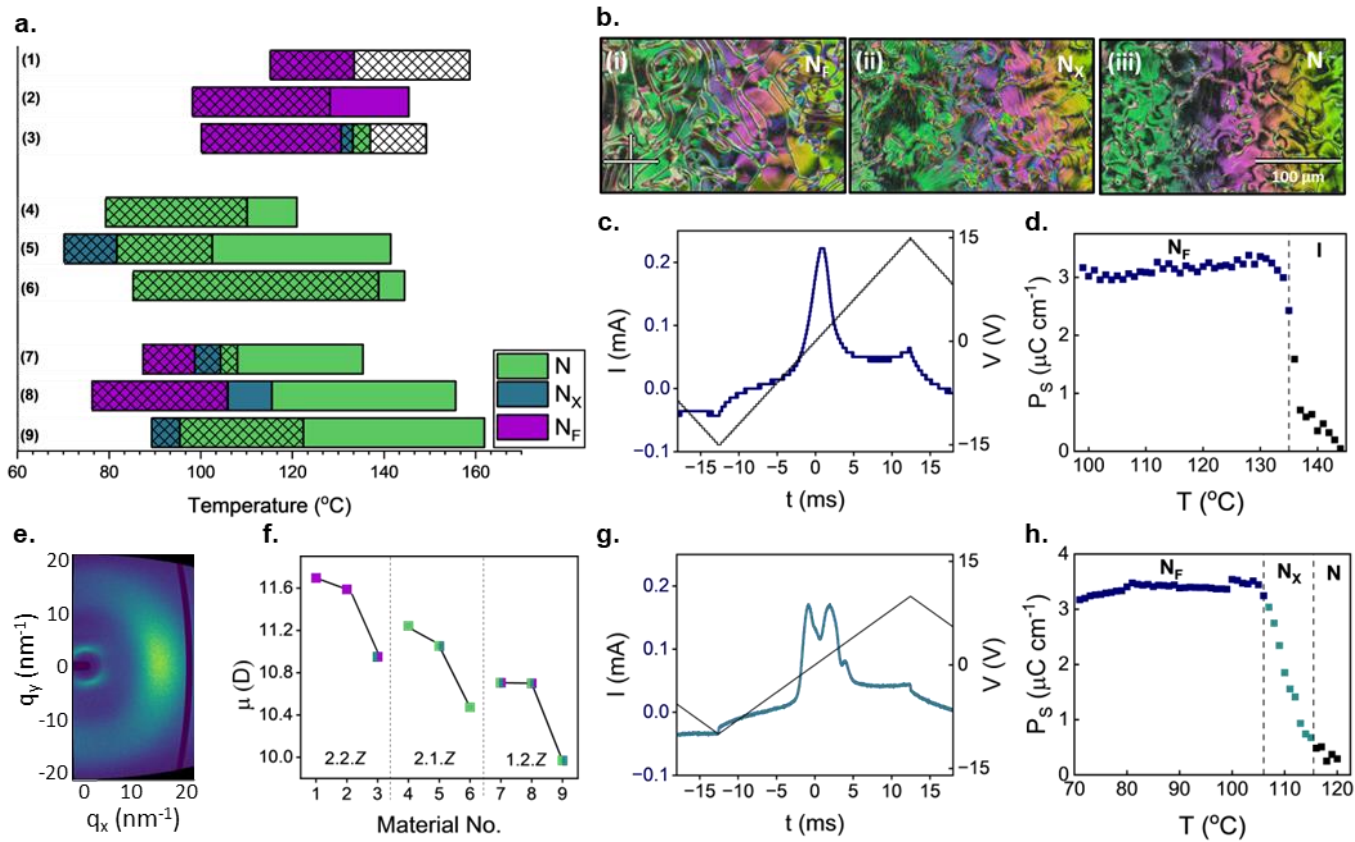

**Figure 3.** (a) the dependence of the transition temperatures on systematic fluorination for materials **1-9**. The hash bar indicates the meting point of the material, any transition within the hashed region are supercooled below the melting point; (b) POM micrographs depicting the (i) $N_F$, (ii) $N_X$ and (iii) N phases observed for **2.2.0 (3)** at 137, 134, and 131 °C, respectively. Images were taken of a thin sample sandwiched between untreated glass slides; (c) Current response trace measured for **2.2.2 (1)** measured at 20 Hz in the $N_F$ phase at 105 °C; (d) temperature dependence of spontaneous polarization ($P_S$) measured for **2.2.2 (1)**; (e) 2D X-ray scattering pattern obtained for **1.2.1 (8)** at 98°C in the $N_F$ phase showing the nematic-like ordering of molecules ; (f) the dependence on the magnitude of the longitudinal molecular dipole moment ($\mu$) on systematic fluorination. The colour of each data point indicates the phase sequence exhibited by the homologue; (g) Current response trace measured for **1.2.1 (8)** measured at 20 Hz in the $N_X$ phase at 110 °C; and (h) temperature dependence of spontaneous polarization ($P_S$) measured for **1.2.1 (8)**.

Further removal of a fluorine substituent from the ***Z*-ring** affords **2·2·0 (3)**. For **2·2·0**, the $N_F$ phase is proceeded by a paraelectric N and subsequent anti-ferroelectric nematic ($N_X$), sometimes referred to as the $N_S$ [29] or $SmZ_A$ [30], phase. The $N_X$ phase was identified by the appearance of a distorted banded texture by POM (**Figure 1b(ii)**) and a double peak in the current response either side of voltage polarity reversal under an applied electrical field (**Figure 1g**). The spontaneous polarisation also saturates almost immediately at the $N_X$-$N_F$ phase transition rather than showing a continuous increase towards the saturation value of $P_S$ which is more generally observed for ferroelectric nematogens (for example see **Figure 1h**) [2,31–34]. Despite the increase in $T_{N_F\text{-}I}$ observed when removing a single fluorine substituent, removal of a further fluorine (**2·2·0**) results in a significant decrease in the transition temperatures associated with polar order, in this case the $N_X$-N transition, compared to the most fluorinated homologue. This modification also decreases the value of µ.

Decreasing the number of F substituents on either the *X*- or *Y*-rings yields two pairs of isomeric structures, **2·1·Z** (4-6) and **1·2·Z** (7-9). Although the **2·1·Z** materials possess notably higher molecular electrical dipole moments (**Figure 1f**), the three **1·2·Z** homologues exhibit a greater number of polar LC phases with their associated transitions to polar order occurring at higher temperatures (i.e. higher values of $T_{NX-N}$). This is perhaps a surprising observation which reinforces the emerging observation that, beyond molecules possessing a sufficient molecular electrical dipole moment such that a polar nematic phase may form, practically the magnitude of µ does not appear to impact the thermal stability of polar nematic phases. Following on from this, when considering the fluorination pattern of the *Z*-ring in materials **4-9**, the values of $T_{N-I}$ behave similarly to **10-27** discussed above whereby increasing the number of F atoms leads to a decrease in the nematic to isotropic transition temperatures. Despite this expected behaviour in $T_{N-I}$, we still observed that homologues with *Z*= 1 have more stable polar phases, evidenced by their higher $N_X$-N transition temperatures ($T_{NX-N}$). When taken together, these two rather surprising observations indicate that the molecular origins of polar nematic phase behaviour are clearly different from those describing the formation of the conventional nematic phase. A complete model describing the formation of polar nematic phases would clearly be highly complex, more so than one describing the formation of conventional nematic materials, and such a model would clearly have to go beyond the basic idea of molecules possessing large molecular dipole moments.

Madhusudana proposed a model in which polar order is suggested to arise from to side-to-side electrostatic interactions between molecules [35]. For the conventional, apolar nematic phase; molecules tend to preferentially adopt anti-parallel conformations relative to their closest neighbours as this helps minimise the dipolar energy of the system [36,37]. Madhusudana suggests that it is possible for molecules to adopt parallel orientations if the electron static potential (ESP) along the long molecular axis oscillates between areas of positive and negative potential as this results in attractive interactions between parallel neighbours [35]. The model has been applied to a variety of known ferroelectric nematogens [38–41], to explain changes in polar LC phase behaviour. Whilst a model based solely on surface charge interactions alone likely cannot completely account for the formation of the $N_F$ phase, an opinion also supported by considering how these electrostatic interactions actually contribute to the free energy of these systems [25,26], electrostatic interactions are likely a significant factor in stabilising longitudinally polar LC phases and are intrinsically linked the molecular structure of these polar LCs.

The systematic approach to selective fluorination undertaken in this work provides us with the unique opportunity to apply the model proposed by Madhusudana to an entire series of homologues where we have a number of homologues exhibiting both polar and apolar nematic phases. To do this, we compute the 3D molecular ESP isosurface (at the DFT:B3LYP-GD3BJ/cc-pVTZ level [42–45]) and radially average the ESP at an electron density isovalue of 0.0004 as a function of the long molecular axis for all **1-27**, allowing the longitudinal ESP surface to be visualised in 1D space (**Figure 2a**). The resultant 1D ESP plots provide insight into potential, favourable lateral interactions which stabilise the polar nematic phase behaviour observed for a select number of **1-27**. We provide further complete 3D ESP surfaces and the resulting 1D reduced data in the ESI (**Figures S, S4 and S5**) as well as further details of this method.

Inspection of these plots for homologues exhibiting polar phase behaviour (for example **2·2·Z**, **Figure 4b -** Purple) and those who exhibit solely conventional nematic behaviour (**2·0·Z**, **Figure 4c** – Green) reveal stark differences in the longitudinal surface charge density across the biphenyl structure. For the three **2·2·Z** homologues, the charge density oscillates almost sinusoidally across the biphenyl structure, with only small changes in the amplitude of the oscillations. In contrast, the variation in charge density across the **2·0·Z** homologues are more pronounced, lacking a clear oscillatory structure[28].We stress that although the radially averaged ESP of the biphenyl region is overall always positive, regardless of fluorination pattern, appended fluorine atoms tend to induce regions of more negative ESP - leading to more favourable, lateral interactions between parallel molecules. The uniformity of the oscillations for 2.2.Z, 1.2.Z (and to a lesser extent 2.1.*Z*) are indicative of the spatial uniformity of these positive and negative regions of the 3D ESP surface where the regions of positive and negative potential all correspond to regions of similar size, something not observed for homologues **10-27** which contain fewer F atoms (**Figure S6**).

Probing more deeply, when considering the spatial uniformity of the oppositely charged regions on the ESP surface, the greater electronegativity of the nitrile moiety present on the *X*-ring appears to negate the effect of removing a fluorine atom (**Figure 4c (i)**) whereas removal of an F atom from the *Y*-ring results in a less uniform oscillatory structure of surface charge (**Figure 4c (ii)**) and thus correspondingly less stable polar mesophases for the **2.1.*Z*** (**4-6**) molecules vs the **1.2.*Z*** (**7-9**) set despite the former having larger longitudinal molecular dipole moments. Reducing fluorination of the *Z*-ring has a much smaller effect on the structure of the 1D ESP and so fluorination of the *Z*-ring has a much smaller impact on the thermal stability of polar nematic phases though we do note that homologues with the ***X.Y.*1** fluorination patterns consistently have slightly higher polar-apolar transition temperatures (I.e. $T_{NF-I}$ and $T_{NX-N}$). Inspection of the 3D ESP isosurface shows that it is actually the ***X.Y.*1** homologues (for example **Figure 4d [top]**) that have the most spatially uniform ESP as the appended fluorine atom matches to the carbonyl atom of the ester group. Adding or removing fluorine (for example **Figure 4d [bottom]**) distorts the uniformity slightly leading to the destabilisation of polar mesophases for those homologues and hence decreases the stability of the polar mesophases.

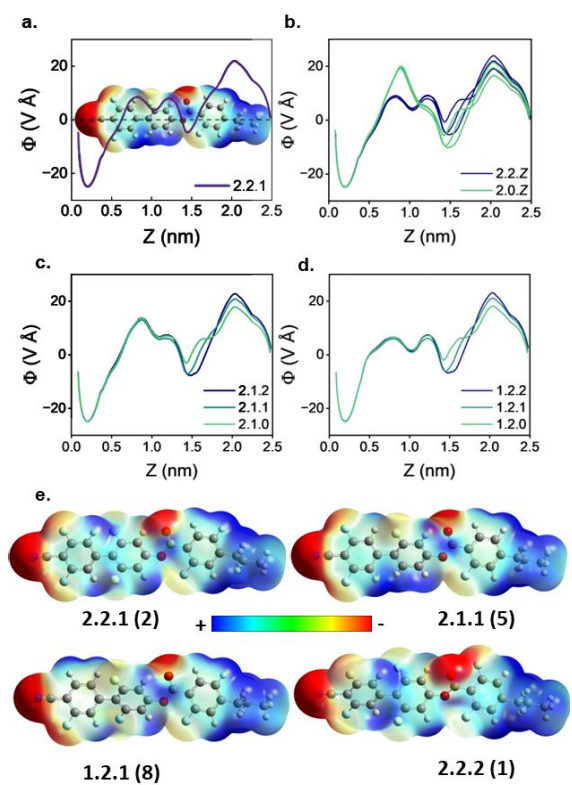

**Figure 4.** **(a)** 3D ESP surface and the resulting 1D longitudinal charge density wave calculated for 2.1.1; (b) the 1D longitudinal charge density waves calculated for the three 2·2·Z and 2·0·Z homologues showing the more uniform oscillatory structure of the charge density wave for the 2·2·Z homologues which results the formation of the polar nematic phases. The greater amplitude of the charge density wave for 2·0·Z promotes anti-parallel associations between the molecules resulting in the solely conventional nematic behaviour observed experimentally; a comparison of the 1D longitudinal charge density waves of the **(c) 2.1.Z (4-6)** and **(d) 1.2.Z (7-9)** (*ii*) homologues, indicating the more oscillatory structure of the latter structures which exhibit more polar nematic phases despite having lower values of μ; and (**e**) 3D ESP surfaces for **2.2.1 (2)** (top left), **2.1.1 (5)** (top right), **1.2.1 (8)** (bottom left) and **2.2.2 (1)** (bottom right). For *Z*= 1, the 3D ESP surface is more spatially uniform due to the position of the appended F atom complimenting the position of the oxygen atom of the ester carbonyl.

Examination of the bimolecular potential energy surface with electronic structure calculations is a logical extension of this simple model. This comprises a rigid bimolecular potential energy scan (PES), beginning from a DFT optimised geometry, in which the position of the second molecule is translated over the x/y/z dimensions (**Figure 5a**). To simplify these calculations, we calculated only the limiting cases of two molecules in a parallel and antiparallel orientation. For each set of translation vectors we obtain the counterpoise corrected complexation energy (at the DFT:B3LYP-GD3BJ/cc-pVTZ level in Gaussian G16 [42–45]). The translation vectors and complexation energies are then used to produce a bimolecular PES (**Figure 5b** for **2·2·2 (1)**). In the antiparallel configuration, repulsive regions are observed that arise from the close proximity of like charges which are absent for the parallel configuration. The size and depth of the repulsive region in the antiparallel configuration is dependent on the degree of fluorination at the nitrile terminus of the molecule. Put another way, the preference for polar order arises, at least in part, from the enthalpic cost of antiparallel packing of such polar rod-like molecules.

Whilst these calculations do give a better understanding of the lateral interaction between **1-27**, changes in the fluorination pattern between homologues affects the specific preferred pairing modes in ways that are difficult to infer from the rigid scans presented here. To that end, and given that both parallel and antiparallel potential energy surfaces have minima with large negative complexation energies, we elected to refine our calculations by extracting five discrete minima for each compound (in both parallel and antiparallel orientations) which we then perform optimisation (at the B3LYP-GD3BJ/cc-pVTZ level). The resultant interaction region indicator (IRI) [46] isosurface allows for the visualisation of the non-covalent interactions between pairs of molecules (**Figure 5c** for **2·2·2 (1)**). In the case of the molecules presented here, the dominant interaction is offset π-π stacking of the biphenyl units with a small contribution arising from the *Z*-ring. Although for none of **1-27** does the global minima in complexation energy for the parallel packed molecules become lower than antiparallel, parallel packing results in multiple positions of relatively comparable energy while antiparallel packing results in only a singular region of highly negative complexation energy as well as regions of repulsive positive complexation energy. Notably, increasing the number of appended fluorine atoms does result in the global minima for each packing mode being considerably closer in energy, particularly for the molecules showing polar phases. This may result in a situation where the increased entropy of multiple possible complexation positions counteracts the slightly increased enthalpic cost of not existing in the global minima, resulting in an overall reduced bulk free energy of parallel arrangement of the molecules.

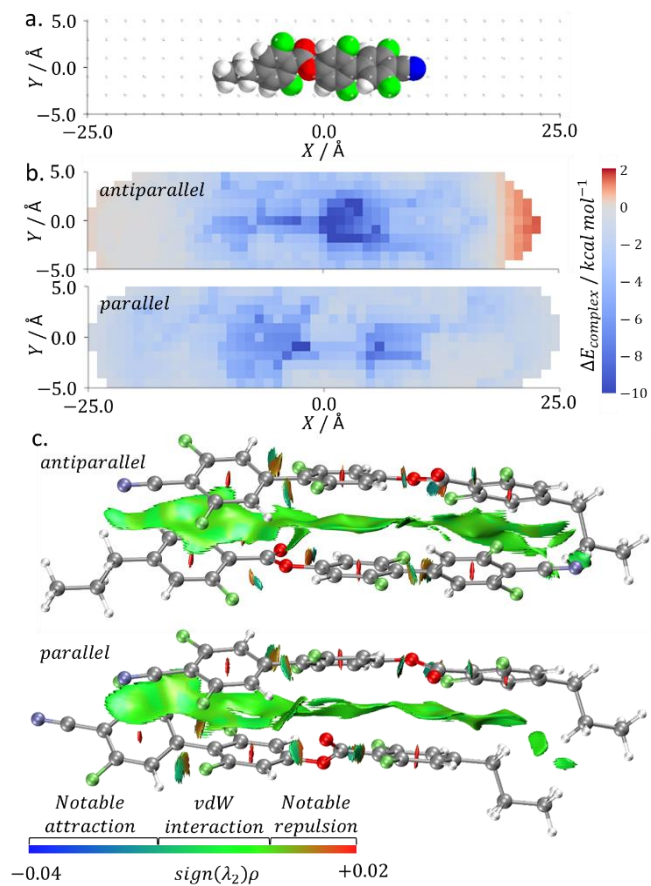

**Figure 5.** (a) Molecular structure of **2·2·2** (**27**) with the grid of translation vectors used in the bimolecular PES shown as points, (b) rigid bimolecular potential energy surfaces for two molecules of **2·2·2** (**27**) in (anti)parallel orientation; a diverging colourscale with midpoint at zero is used to highlight attractive (blue) and repulsive (red) regions, (c) Interaction Region Indicator (IRI) isosurface (isovalue 1.0) for the global energy minimum of the antiparallel and parallel forms of **2·2·2** (**27**), identified *via* the bimolecular PES; the colourbar indicates the different types of noncovalent interactions present on the isosurface.

## Conclusions

In summary, against current thinking, we have shown that maximal fluorination of these materials does not necessarily result in maximal polar mesophase stability, with specific fluorination patterns being preferred. This also highlights that the magnitude of the longitudinal dipole moment is not the most important metric for predicting polar phase behaviour even in chemically similar compounds. We have complimented our synthetic efforts with a series of computational methodologies which provide insight into the molecular origins of polar nematic phase behaviour by probing the lateral interactions between molecules necessary for these phases to form. We show that rather than considering the electrostatic interactions as 1D rod like objects, consideration must be given to the resulting 3D ESP of the molecule. Moreover, evaluation of the bimolecular potential energy surfaces, coupled with the interaction region indicator, shows the dominant mode of interaction between molecules to be offset π-π stacking rather than the often-quoted dipole-dipole interactions.

## Data availability

The data associated with this paper are openly available from the University of Leeds Data Repository at https://doi.org/10.5518/1573.

## Author Contribution Statement

CJG and JH contributed equally to this work. CJG and RJM performed chemical synthesis; JH performed X-ray scattering experiments, DSC and, applied field studies; CJG and JH performed microscopy studies, JH performed and evaluated electronic structure calculations; RJM wrote the bimolecular potential energy surface code and secured funding. The manuscript was written, reviewed, and edited with contributions from all authors.

## Competing interests

Authors declare that they have no competing interests.

## Acknowledgements

RJM thanks UKRI for funding via a Future Leaders Fellowship, grant number MR/W006391/1, and the University of Leeds for funding via a University Academic Fellowship. The SAXS/WAXS system used in this work was funded by EPSRC via grant number EP/X0348011. R.J.M. gratefully acknowledge support from Merck KGaA. Computational work was performed on ARC3 and ARC4, part of the high-performance computing facilities at the University of Leeds.

# Systematic Fluorination is a Powerful Design Strategy Towards Fluid Molecular Ferroelectrics

## Supplemental Information

Calum J. Gibb*[1], Jordan Hobbs [2], Richard. J. Mandle [1,2]

[1]School of Chemistry, University of Leeds, Leeds, UK, LS2 9JT
[2]School of Physics and Astronomy, University of Leeds, Leeds, UK, LS2 9JT

*c.j.gibb@leeds.ac.uk

**Contents**

1 **Supplementary Methods**
    1.1 Chemical synthesis
    1.2 Chemical characterisation methods
    1.3 Mesophase characterisation
    1.4 X-ray scattering
    1.5 Measurements of spontaneous polarization ($P_S$)
    1.6 DFT calculations

2 **Supplemental results**

3 **Organic Synthesis**
    3.1 Synthesis of 4-(2 fluoro-4-hydroxyphenyl)-2,6-difluorobenzonitrile.
    3.2 4-(2,6-difluoro-4-hydroxyphenyl)-2-fluorobenzonitrile
    3.3 Synthesis of materials **1-27**
    3.4 Example structural characterisation

4 **Supplemental References**

# 1 Supplementary Methods

## 1.1 Chemical Synthesis

Chemicals were purchased from commercial suppliers (Fluorochem, Merck, Apollo Scientific) and used as received. Solvents were purchased from Merck and used without further purification. Reactions were performed in standard laboratory glassware at ambient temperature and atmosphere and were monitored by TLC with an appropriate eluent and visualised with 254 nm or 365 nm light. Chromatographic purification was performed using a Combiflash NextGen 300+ System (Teledyne Isco) with a silica gel stationary phase and a hexane/ethyl acetate gradient as the mobile phase, with detection made in the 200-800 nm range. Chromatographed materials subjected to re-crystallisation from an appropriate solvent system.

## 1.2 Chemical Characterisation Methods

NMR was performed using a Bruker Avance III HDNMR spectrometer operating at 400 MHz, 100.5 MHz or 376.4 MHz ($^1$H, $^{13}$C{$^1$H} and $^{19}$F, respectively). Unless otherwise stated, spectra were acquired as solutions in deuterated chloroform, coupling constants are quoted in Hz, and chemical shifts are quoted in ppm.

## 1.3 Mesophase Characterisation

Transition temperatures and measurement of associated latent heats were measured by differential scanning calorimetry (DSC) using a TA instruments Q2000 heat flux calorimeter with a liquid nitrogen cooling system for temperature control. Between 3-8 mg of sample was placed into T-zero aluminium DSC pans and then sealed. Samples were measured under a nitrogen atmosphere with 10 °C min$^{-1}$ heating and cooling rates. The transition temperatures and enthalpy values reported are averages obtained for duplicate runs. In general LC phase transition temperatures are measured on cooling from the onset of the transition while melt temperatures were measured on heating to avoid crystallization loops that can occur on cooling. Phase identification by polarised optical microscopy (POM) was performed using a Leica DM 2700 P polarised optical microscope equipped with a Linkam TMS 92 heating stage. Samples were studied sandwiched between two untreated glass coverslips.

## 1.4 X-ray Scattering

X-ray scattering measurements, both small angle (SAXS) and wide angle (WAXS) where recorded using an Anton Paar SAXSpoint 5.0 beamline machine. This was equipped with a primux 100 Cu X-ray source with a 2D EIGER2 R detector. The X-rays had a wavelength of 0.154 nm. Samples were filled into thin-walled quartz capillaries 1 mm thick. Temperature was controlled using an Anton Paar heated sampler with a range of -10 ℃ to 107 ℃ and the samples held in a chamber with an atmospheric pressure of <1 mBar. Samples were held at 107 ℃ to allow for temperature equilibration across the sample and then slowly cooled while stopping to record the 2D scattering patterns. The 2D patterns are then radially integrated to obtain 1D patterns.

## 1.5 Measurement of Spontaneous Polarization (P$_S$)

Spontaneous polarisation measurements are undertaken using the current reversal technique [1,2]. Triangular waveform AC voltages are applied to the sample cells with an Agilent 33220A signal generator (Keysight Technologies), and the resulting current outflow is passed through a current-to-voltage amplifier and recorded on a RIGOL DHO4204 high-resolution oscilloscope (Telonic Instruments Ltd, UK). Heating and cooling of the samples during these measurements is achieved with an Instec HCS402 hot stage controlled to 10 mK stability by an Instec mK1000 temperature controller. The LC samples are held in 4µm thick cells with no alignment layer, supplied by Instec. The measurements consist of cooling the sample at a rate of 1 Kmin$^{-1}$ and applying a set voltage at a frequency of 10 Hz. The voltage was set such that it would saturate the measured $P_S$ and was determined before final data collection.

There are three contributions to the measured current trace: accumulation of charge in the cell ($I_c$), ion flow ($I_i$), and the current flow due to polarisation reversal ($I_p$). To obtain a $P_S$ value, we extract the latter, which manifests as one or multiple peaks in the current flow, and integrate as:

$$P_S = \int \frac{I_p}{2A} dt \quad \textbf{(2)}$$

where A is the active electrode area of the sample cell. For the N, $N_X$ and, to a lesser extent, the $N_F$ phase, significant amounts of ion flow is present. For materials that showed a paraelectric N phase followed by the anti-ferroelectric $N_X$ phases, the N phase always showed some pre-transitional polarisation as well as the significant ion flow mentioned previously. The $P_S$ of the $N_X$ phases was obtained by integrating the peak least affected by ion flow and then doubled to get the total area under both peaks [3].

### 1.6 DFT Calculations

Electronic structure calculations were performed using Gaussian G16 revision C.02 [4] and with a B3LYP-GD3BJ/cc-pVTZ [5–8] basis set. Obtained structures were verified as a minimum from frequency calculations. Electrostatic potential (ESP) surfaces were calculated by using the *formchk* and *cubegen* utilities. Both the electron density and ESP cube files were calculated using "fine" data resolution. The 3D ESP surface is displayed at an electron density iso-surface of 0.0004.

The 3D data was reduced into 1D through the following steps. The electron density and ESP cube files are structured such that the long molecule axis of the molecule is centred along the z-axis of the data in each cube file. Each step in the z-axis is taken as a single plane through the molecule at that point. An iso-contour through the electron density cube file is found at some isovalue (here 0.0004 as used to mimic the 3D surfaces). The values of the ESP data that then fall on this iso-contour route are then found. These values reflect the 3D surface visualised in figure S6 exactly. We assume free rotation around the long molecule axis and so average the entire ESP data that falls along the iso-contour. This gives the average ESP value that a neighbouring molecule will "feel" for timescales longer that those of rotation around the long axis.

A further step of rescaling the values obtained by the length of the contour allows to account for the fact that at the molecular extremes the values are distorted by the reduction in molecular volume. This final step effectively gives the ESP as electric flux i.e. the strength of the electric field due to the molecular dipole through the contour.

## 2    Supplementary Results

**Table S1.** Phase sequences of **1-27** and their associated transition temperatures (°C) The fluorination pattern value indicates the number of fluorine atoms of the 4-(per)flurobenzoic acid (indicated on the structure below). **24** degraded before $T_{NI}$ was identified

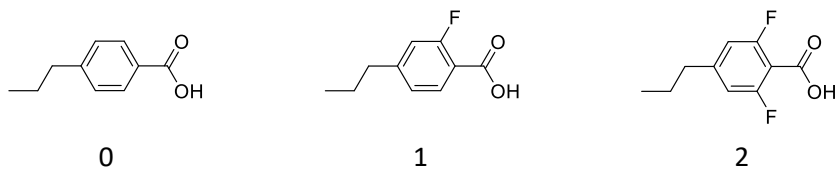

| Compound | 1-nitrile-n-n'-(per)flurobiphenyl-4'-phenol | Fluorination pattern | Phase Sequence and associated transition temperatures / °C |
|---|---|---|---|
| 1 | | 2 | K 158.7 $N_F$ 133.5 Iso |
| 2 | | 1 | K 128.2 $N_F$ 145.4 Iso |
| 3 | | 0 | K 149.2 $N_F$ 130.6 $N_X$ 133.2 N 137.0 Iso |
| 4 | | 2 | K 138.8 N 121.0 Iso |
| 5 | | 1 | K 102.5 $N_X$ 81.7 N 141.5 Iso |
| 6 | | 0 | K 110.1 N 144.5 Iso |
| 7 | | 2 | K 108.0 $N_F$ 98.7 $N_X$ 104.3 N 135.4 Iso |
| 8 | | 1 | K 105.9 $N_F$ 105.8 $N_X$ 115.5 N 155.6 Iso |
| 9 | | 0 | K 122.4 $N_X$ 95.4 N 161.9 Iso |
| 10 | | 2 | K 119.9 N 153.8 Iso |
| 11 | | 1 | K 111.9 N 177.1 Iso |
| 12 | | 0 | K 133.4 N 187.7 Iso |
| 13 | | 2 | K 99.0 N 150.9 Iso |
| 14 | | 1 | K 121.2 N 181.8 Iso |
| 15 | | 0 | K 113.7 N 187.9 Iso |
| 16 | | 2 | K 124.5 N 110.8 Iso |
| 17 | | 1 | K 104.9 N 151.5 Iso |
| 18 | | 0 | K 98.6 N 149.3 Iso |
| 19 | | 2 | K 105.3 N 165.1 Iso |
| 20 | | 1 | K 84.6 N 200.2 Iso |
| 21 | | 0 | K 87.6 N 215.3 Iso |
| 22 | | 2 | K 91.1 N 176.6 Iso |
| 23 | | 1 | K 108.2 N 204.9 Iso |
| 24 | | 0 | K 112.4 N >220.0 Iso |
| 25 | | 2 | K 109.9 N 205.1 Iso |
| 26 | | 1 | K 100.7 SmA 71.4 N 241.8 Iso |
| 27 | | 0 | K 100.6 N 255.6 Iso |

**Table S2.** Phase sequences of **1-27** and their associated enthalpy changes (kJ/mol) The fluorination pattern value indicates the number of fluorine atoms of the 4-(per)flurobenzoic acid (indicated on the structure below).

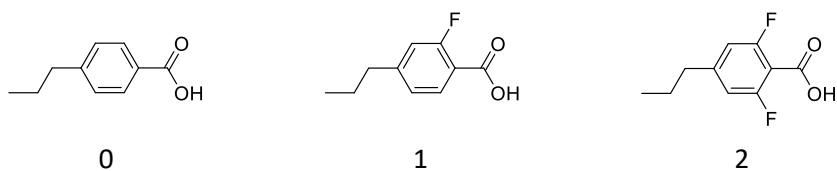

| Cpd. No. | Biphenyl-type | Fluorination pattern | $\Delta H_{fus}$ (kJ/mol) | $\Delta H_{N-SmA}$ (kJ/mol) | $\Delta H_{N-N_F}$ (kJ/mol) | $\Delta H_{N-N_X}$ (kJ/mol) | $\Delta H_{I-N_F/N}$ (kJ/mol) |
|---|---|---|---|---|---|---|---|
| 1 | 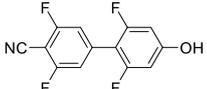 | 2 | 38.8 | - | - | - | 4.7 |
| 2 |  | 1 | 25.6 | - | - | - | 4.7 |
| 3 |  | 0 | 35.5 | - | 0.6 | 0.03 | 1.2 |
| 4 | 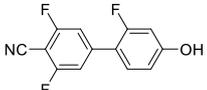 | 2 | 35.9 | - | - | - | 0.6 |
| 5 |  | 1 | 25.3 | - | - | 0.02 | 0.4 |
| 6 |  | 0 | 27.8 | - | - | - | 0.4 |
| 7 | 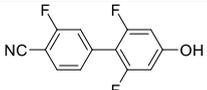 | 2 | 36.0 | - | 0.3 | 0.01 | 0.9 |
| 8 |  | 1 | 38.1 | - | 0.3 | 0.01 | 0.7 |
| 9 |  | 0 | 30.9 | - | - | 0.01 | 1.0 |
| 10 | 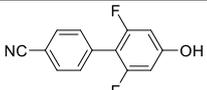 | 2 | 29.3 | - | - | - | 1.2 |
| 11 |  | 1 | 24.3 | - | - | - | 1.4 |
| 12 |  | 0 | 31.6 | - | - | - | 1.4 |
| 13 | 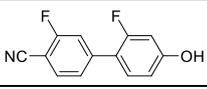 | 2 | 37.5 | - | - | - | 0.6 |
| 14 |  | 1 | 17.3 | - | - | - | 0.7 |
| 15 |  | 0 | 23.3 | - | - | - | 0.6 |
| 16 | 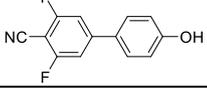 | 2 | 30.0 | - | - | - | 0.2 |
| 17 |  | 1 | 19.9 | - | - | - | 0.3 |
| 18 |  | 0 | 27.2 | - | - | - | 0.5 |
| 19 | 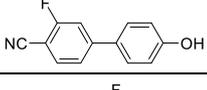 | 2 | 33.4 | - | - | - | 0.5 |
| 20 |  | 1 | 25.3 | - | - | - | 0.6 |
| 21 |  | 0 | 17.7 | - | - | - | 0.8 |
| 22 | 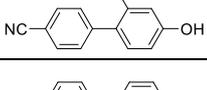 | 2 | 23.4 | - | - | - | 1.1 |
| 23 |  | 1 | 26.1 | - | - | - | 1.1 |
| 24 |  | 0 | 30.1 | - | - | - | - |
| 25 | 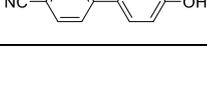 | 2 | 30.6 | - | - | - | 0.7 |
| 26 |  | 1 | 27.0 | 0.08 | - | - | 0.8 |
| 27 |  | 0 | 25.4 | - | - | - | 1.1 |

**Table S3.** DFT parameters calculated at the DFT:B3LYP-GD3BJ/cc-pVTZ level of theory for compounds **1-27**.

| Cpd. No. | Biphenyl-type | Fluorination pattern | Dipole Moment / D | Dipole Angle / ° | Length / nm | Width / nm | Aspect Ratio |
|---|---|---|---|---|---|---|---|
| 1 |  | 2 | 11.70 | 10.2 | 2.028 | 0.5.09 | 3.99 |
| 2 |  | 1 | 11.59 | 3.6 | 2.019 | 0.4.99 | 4.05 |
| 3 |  | 0 | 10.95 | 8.8 | 2.014 | 0.5.01 | 4.02 |
| 4 |  | 2 | 11.24 | 14.0 | 2.026 | 0.4.93 | 4.11 |
| 5 |  | 1 | 11.05 | 7.1 | 2.018 | 0.5.11 | 3.95 |
| 6 |  | 0 | 10.47 | 12.5 | 2.012 | 0.511 | 3.94 |
| 7 |  | 2 | 10.70 | 6.3 | 2.026 | 0.4.99 | 4.06 |
| 8 |  | 1 | 10.70 | 1.2 | 2.019 | 0.4.74 | 4.26 |
| 9 |  | 0 | 9.97 | 4.5 | 2.013 | 0.4.77 | 4.22 |
| 10 |  | 2 | 9.93 | 11.9 | 2.027 | 0.51 | 3.97 |
| 11 |  | 1 | 9.79 | 4.3 | 2.019 | 0.48 | 4.18 |
| 12 |  | 0 | 9.18 | 10.5 | 2.013 | 0.48 | 4.17 |
| 13 |  | 2 | 10.73 | 19.6 | 2.026 | 0.50 | 4.06 |
| 14 |  | 1 | 10.40 | 12.6 | 2.018 | 0.52 | 3.91 |
| 15 |  | 0 | 9.96 | 18.7 | 2.012 | 0.52 | 3.90 |
| 16 |  | 2 | 10.39 | 11.8 | 2.026 | 0.49 | 4.13 |
| 17 |  | 1 | 10.25 | 4.5 | 2.017 | 0.50 | 4.04 |
| 18 |  | 0 | 9.66 | 10.5 | 2.012 | 0.50 | 4.03 |
| 19 |  | 2 | 9.36 | 7.5 | 2.024 | 0.49 | 4.10 |
| 20 |  | 1 | 9.34 | 2.4 | 2.017 | 0.47 | 4.25 |
| 21 |  | 0 | 8.64 | 6.3 | 2.011 | 0.48 | 4.23 |
| 22 |  | 2 | 9.50 | 16.6 | 2.026 | 0.49 | 4.13 |
| 23 |  | 1 | 9.25 | 8.7 | 2.017 | 0.49 | 4.09 |
| 24 |  | 0 | 8.72 | 15.2 | 2.011 | 0.49 | 4.07 |
| 25 |  | 2 | 8.61 | 14.0 | 2.026 | 0.49 | 4.13 |
| 26 |  | 1 | 8.42 | 5.4 | 2.016 | 0.48 | 4.19 |
| 27 |  | 0 | 7.86 | 12.8 | 2.011 | 0.48 | 4.17 |

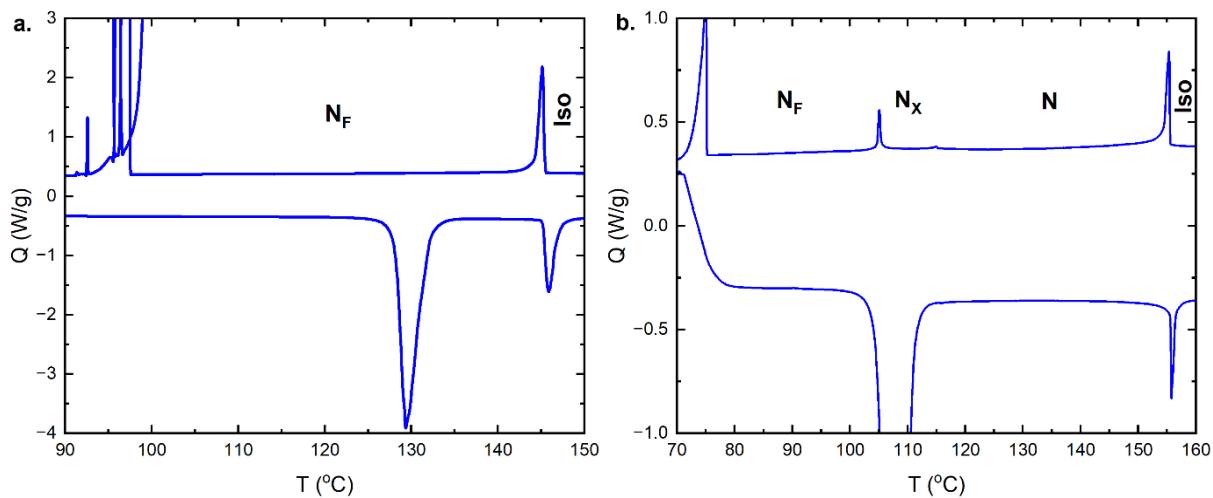

**Figure S1.** Example DSC traces for (a) **(2)** and (b) **(8)**; the exothermic direction is upwards.

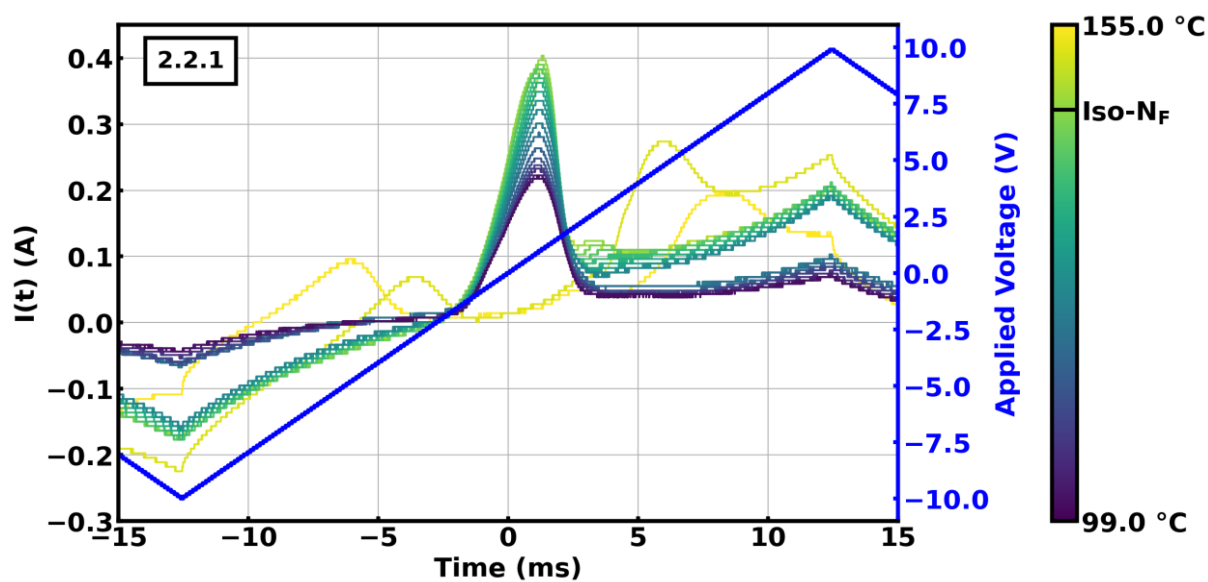

**Figure S2.** Current responses for compound **2** showing the pre-transitional double peaks (yellow data) associated with the field induced phase transition from the isotropic state to the $N_F$.

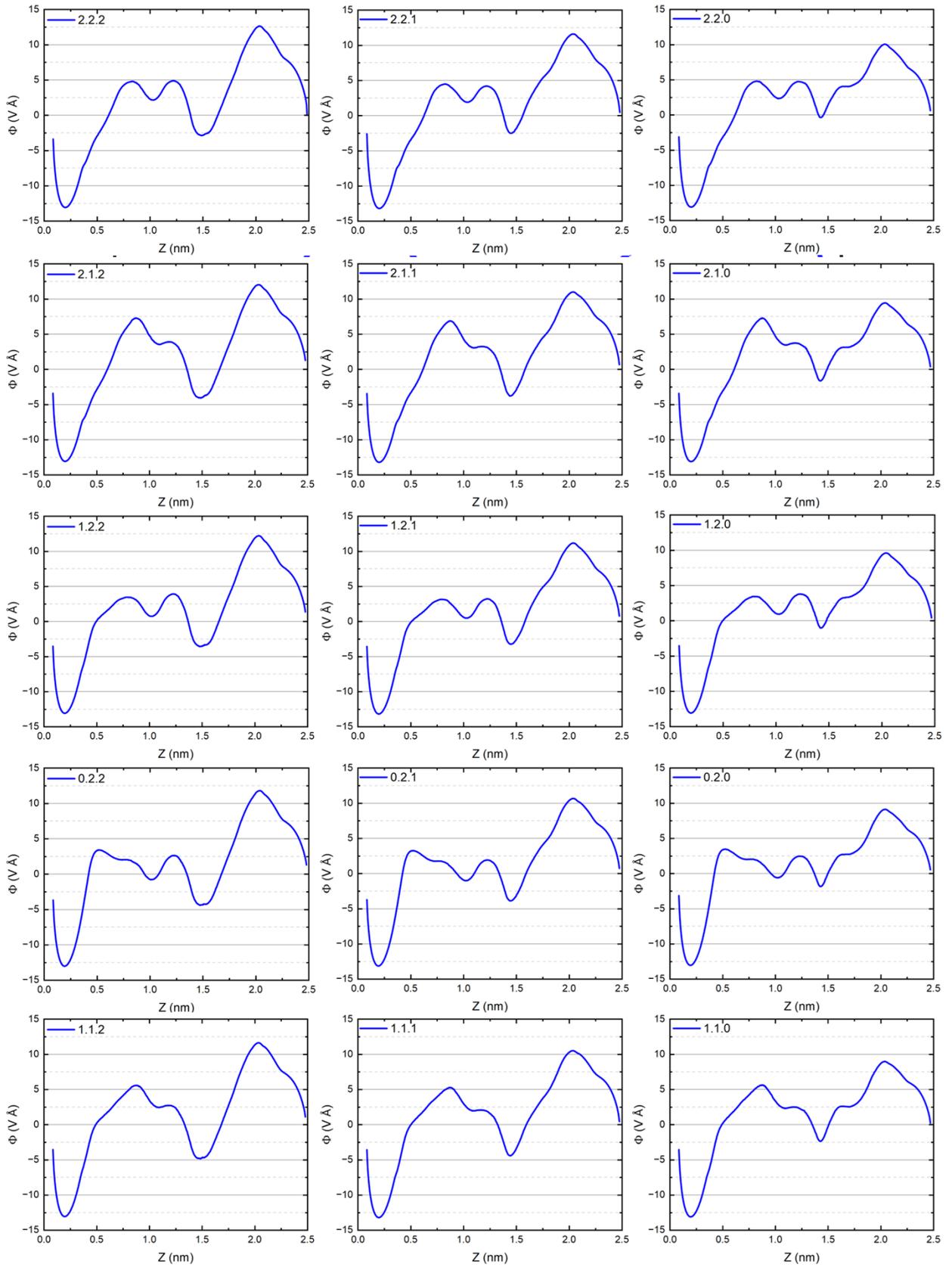

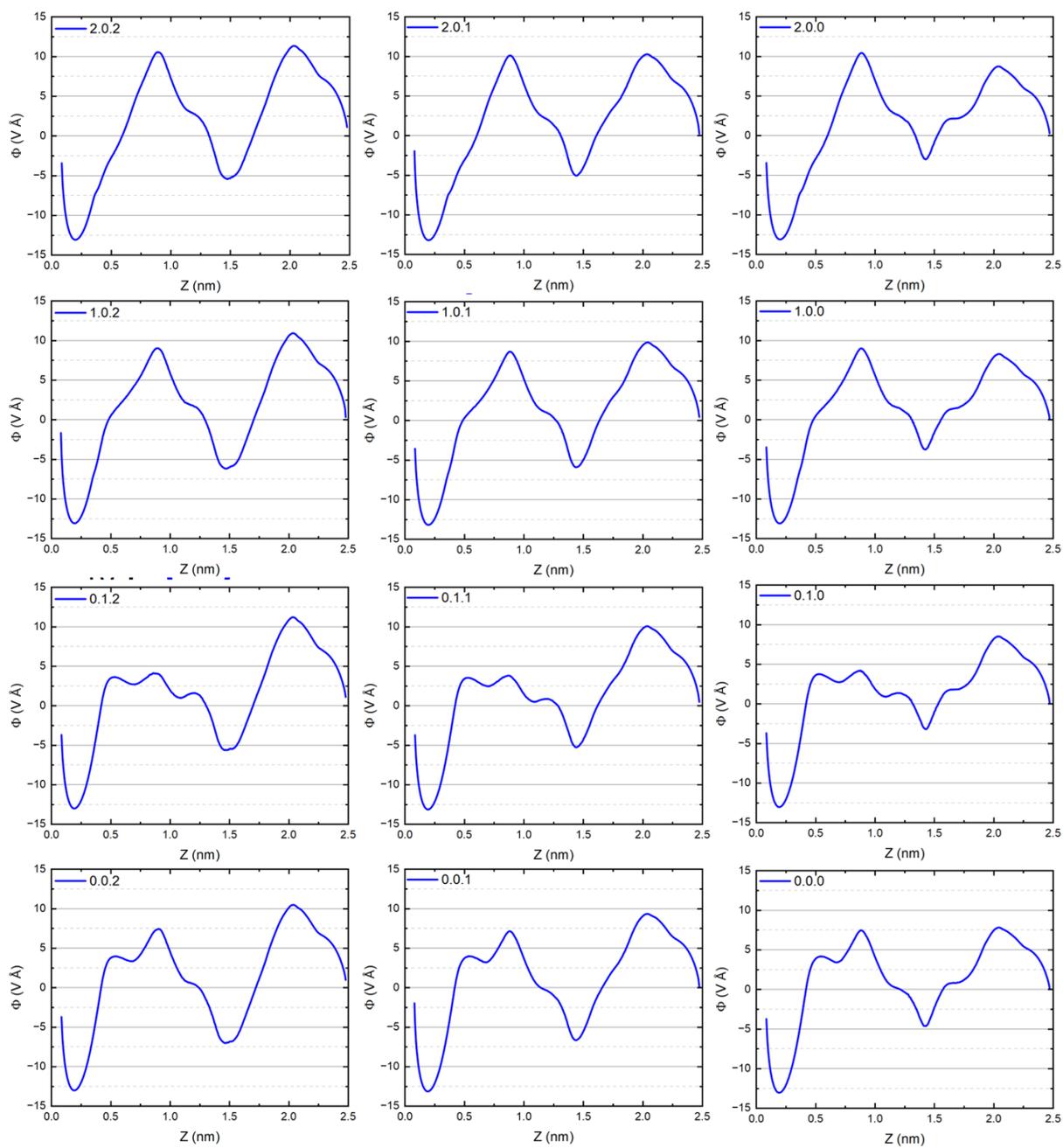

**Figure S3**. 1D reduced scaled ESP data for materials **1-27**.

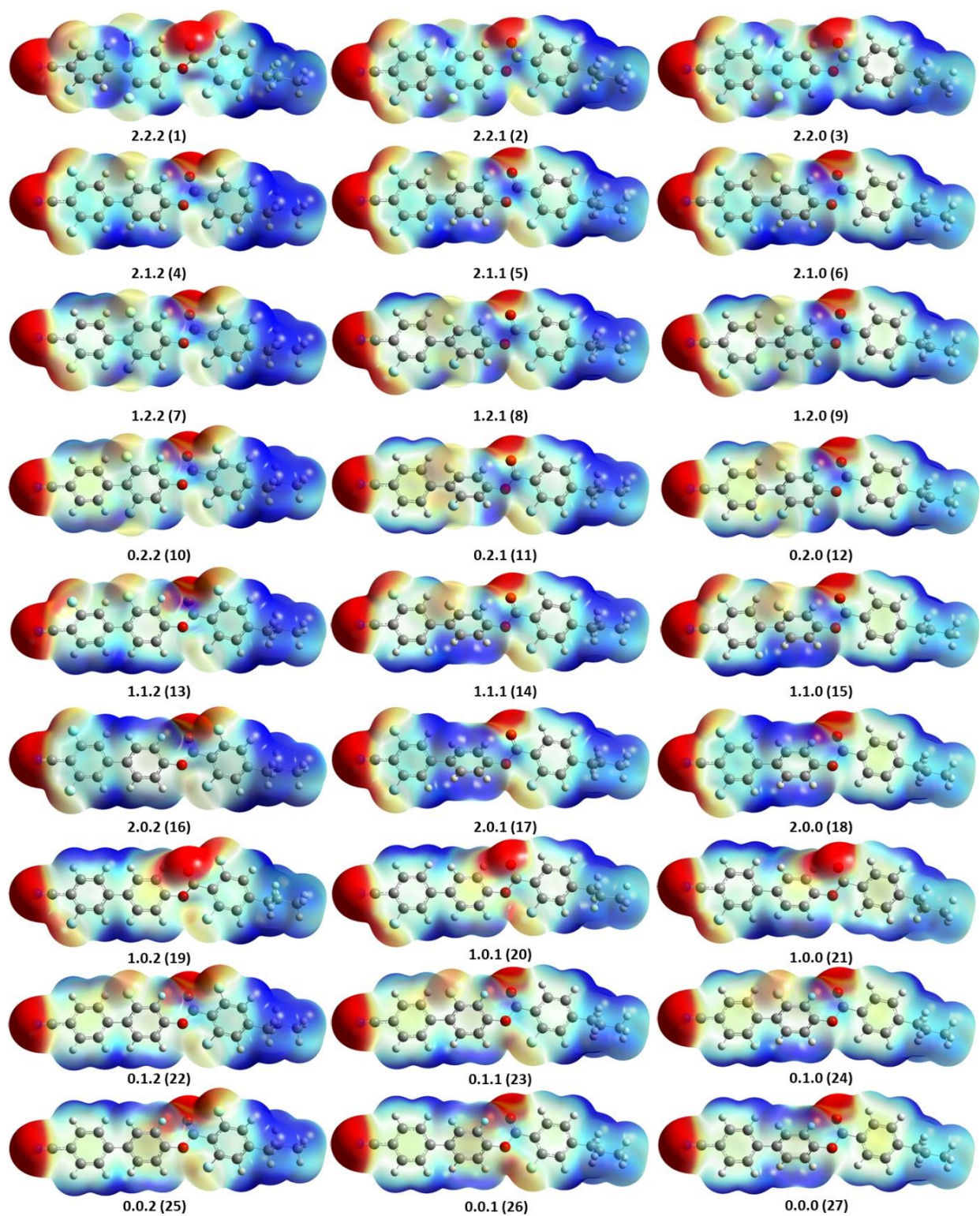

**Figure S4** 3D ESP data for materials **1-27**. Further blue indicates more positive while further red indicates more negative. Obtained for an electron density isovalue of 0.0004

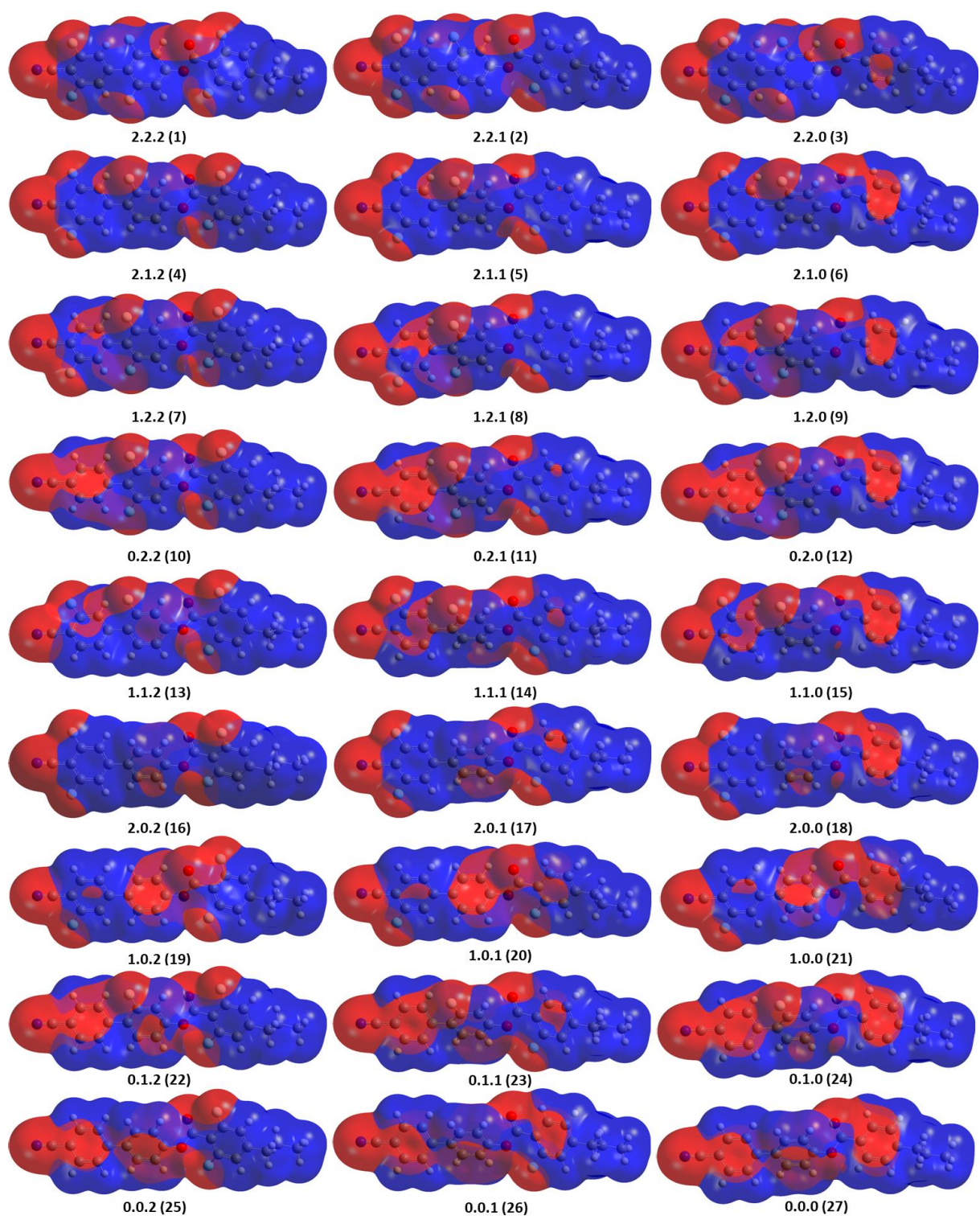

**Figure S5**. 3D ESP data for materials **1-27**. Obtained for an electron density isovalue of 0.0004. Here all negative regions have been block-coloured red while all the positive regions have been block coloured blue.

## 3   Organic synthesis

The total synthesis of materials **1-27** is outlined in **Scheme 1** (main manuscript). The synthesis of the 2-fluoro and 2,6- difluoro-4-propyl benzoic acids is described elsewhere [9], and 4-propyl benzoic acid is an article of commerce. The synthesis of some of the fluoro biphenyl phenols have also been described previously within the literature [10–13], and 4-hydroxy-4′-cyanobiphenyl is also an article of commerce.

### 3.1   Synthesis of 4-(2 fluoro-4-hydroxyphenyl)-2,6-difluorobenzonitrile.

A reaction flask was charged with 4-bromo-2,6-difluorobenzonitrile (8.4 g, 38.5 mmol) and 2-fluoro-4-methoxyphenylboronic acid (7.16 g, 42.3 mmol) which were dissolved in 100 mL of THF and 60 mL of 2M $Na_2CO_{3(aq)}$. The resultant solution was sparged with $N_{2(g)}$ for 20 minutes. In a separate vial, 5 mL of THF was sparged with $N_{2(g)}$ for 15 minutes before $Pd(OAc)_2$ (50 mg) and SPhos (100 mg) were added and stirred for a further 5 minutes. The reaction flask was then heated to 70 °C and the catalyst solution added in one portion. The reaction was monitored by TLC with the completion of the reaction being determined by the complete consumption of the bromo-sub-straight ($R_{f\ prod.}$[DCM] = 0.84). The reaction was then cooled, the aqueous and organic layers separated with the organics being dried over $MgSO_4$. The organics were then passed through a silica plug before the filtrate was concentrated under reduced pressure and the product re-crystallised from hexane.

The resultant product (4.5 g, 45%) was then immediately carried forward, dissolved in DCM (conc. ~1M) under an atmosphere of $N_2$. A solution of $BBr_3$ (1M in DCM, 30 mL, 30 mmol) was then added dropwise to the stirred solution with the progress of the reaction monitored by TLC ($R_{f\ prod.}$[DCM] ≈ 0). Once complete, the reaction mixture was quenched with water, extracted and dried over $MgSO_4$. The reaction solution was concentrated and purified by flash chromatography over silica gel with a gradient of hexane/ethyl acetate using a Combiflash NextGen300+ system using a gradient elution from hexane - ethyl acetate. The product was then recrystallized from toluene as fine white solid.

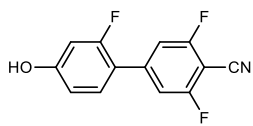

*4-(2 fluoro-4-hydroxyphenyl)-2,6-difluorobenzonitrile*

$R_F$ (DCM): ≈ 0

$^1$H NMR (400 MHz, DMSO) (δ): 10.63 (s, 1H, Ar-O**H**), 7.57 – 7.45 (m, 3H, Ar-**H**)*, 6.80 – 6.71 (m, 2H, Ar-**H**)*.

*overlapping signals

$^{13}$C{$^1$H} NMR (101 MHz, DMSO) (δ): 163.97 (d, J = 5.6 Hz), 161.69 (d, J = 251.0 Hz), 161.42 (dd, J = 9.9, 3.0 Hz), 144.68 (t, J = 10.9 Hz), 131.86, 115.34 (d, J = 11.5 Hz), 113.14 (d, J = 2.1 Hz), 112.43 (dt, J = 20.6, 4.0 Hz), 110.17, 103.79 (d, J = 24.8 Hz), 89.27 (t, J = 19.6 Hz).

¹⁹F NMR (376 MHz, DMSO) (δ): -106.13 (d_apparent, J = 10.5 Hz, 2F, Ar-**F**), -114.75 (t_apparent, J = 11.5 Hz, 1F, Ar-**F**).

## 3.2    Synthesis of 4-(2,6-difluoro-4-hydroxyphenyl)-2-fluorobenzonitrile

A reaction flask was charged with 4-bromo-2-fluorobenzonitrile (8.0 g, 40 mmol) and 2,6-difluoro-4-methoxyphenylboronic acid (8.4 g, 44 mmol) which were dissolved in 125 mL of THF and 20 mL of 2M $K_2CO_{3(aq)}$. The resultant solution was sparged with $N_{2(g)}$ for 20 minutes. In a separate vial, 5 mL of THF was sparged with $N_{2(g)}$ for 15 minutes before Pd(OAc)$_2$ (20 mg) and SPhos (40 mg) were added and stirred for a further 5 minutes. The reaction flask was then heated to 70 °C and the catalyst solution added. The reaction was monitored by TLC with the completion of the reaction being determined by the complete consumption of the bromo-sub-straight (R$_{f\ prod.}$[DCM] = 0.85). The reaction was then cooled, the aqueous and organic layers separated with the organics being dried over MgSO$_4$. The organics were then passed through a silica plug before the filtrate was concentrated under reduced pressure and the product re-crystallised from hexane.

The resultant product (6.5 g, 62%) was then immediately carried forward, dissolved in DCM (conc. ~1M) under an atmosphere of N$_2$. A solution of BBr$_3$ (1M in DCM, 30 mL, 30 mmol) was then added dropwise to the stirred solution with the progress of the reaction monitored by TLC (R$_{f\ prod.}$[DCM] ≈ 0). Once complete, the reaction mixture was quenched with water, extracted and dried over MgSO$_4$. The reaction solution was concentrated and purified by flash chromatography over silica gel with a gradient of hexane/ethyl acetate using a Combiflash NextGen300+ system. The product was then recrystallized from toluene as fine white solid.

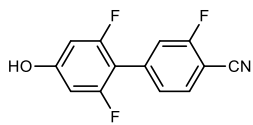

*4-(2,6-difluoro-4-hydroxyphenyl)-2-fluorobenzonitrile*

R$_F$ (DCM): ≈ 0

¹H NMR (400 MHz, DMSO) (δ): 10.78 (s, 1H, Ar-O**H**), 7.99 (dd, J = 8.1, 7.1 Hz, 1H, Ar-**H**), 7.61 (dd, J = 10.6, 1.3 Hz, 1H, Ar-**H**), 7.45 (dd, J = 8.1, 1.5 Hz, 1H, Ar-**H**), 6.74 – 6.50 (m_apparent, 2H, Ar-**H**).

¹³C{¹H} NMR (101 MHz, DMSO) (δ): 163.85, 161.41 (dd, J = 246.0, 8.9 Hz), 161.32, 160.59 (t, J = 15.1 Hz), 137.52 (d, J = 9.2 Hz), 134.14, 127.78, 118.48 (d, J = 20.6 Hz), 114.39, 106.43 (t, J = 18.9 Hz), 100.19 (dd, J = 27.6, 6.4 Hz), 99.70 (d, J = 15.0 Hz).

¹⁹F NMR (376 MHz, DMSO) (δ): -108.58 (dd, J = 10.6, 7.0 Hz, 1F, Ar-**F**), -114.55 (d, J = 10.9 Hz, 2F, Ar-**F**).

### 3.3 Synthesis of materials 1-27.

A round bottomed flask was charged with the appropriate phenol (1 mmol, 1.0 eq), benzoic acid (1.1 mmol, 1.1 eq.), EDC.HCl (1.5 mmol, 1.5 eqv.) and DMAP (~ 2mol%). Dichloromethane was added (conc. ~ 0.1 M) and the suspension stirred until complete consumption of the phenol as judged by TLC. Once complete, the reaction solution was concentrated and purified by flash chromatography over silica gel with a gradient of hexane/DCM using a Combiflash NextGen300+ system. The chromatographed material was dissolved into the minimum quantity of DCM, filtered through a 0.2 micron PTFE filter, concentrated to dryness and finally recrystalised from EtOH.

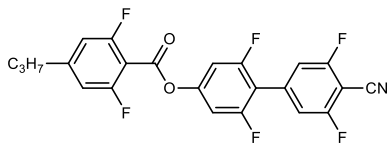

**1** (**2.2.2**): *4'-Cyano-2,3',5',6 tetrafluoro-[1,1' biphenyl]4-yl 2,6 difluoro-4-propyl benzoate*

Yield: (Shiny white solid) 323 mg, 72%

$R_F$ (DCM): 0.81

$^1$H NMR (400 MHz, CDCl$_3$) (δ): 7.22 (d$_{apparent}$, J = 9.0 Hz, 2H, Ar-**H**), 7.06 (ddd, J = 8.5, 1.4, 1.4 Hz, 2H, Ar-**H**), 6.87 (d$_{apparent}$, J = 9.8 Hz, 2H, Ar-**H**), 2.65 (t, J = 7.6 Hz, 2H, Ar-C**H$_2$**-CH$_2$), 1.69 (q, J = 7.5 Hz, 2H, CH$_2$-C**H$_2$**-CH$_3$), 0.98 (t, J = 7.3 Hz, 3H, CH$_2$-C**H$_3$**).

$^{13}$C{$^1$H} NMR (101 MHz, CDCl$_3$) (δ): 162.76 (dd, J = 261.6, 5.1 Hz), 160.83 (dd, J = 174.5, 5.9 Hz), 159.16 (dd, J = 168.7, 5.9 Hz), 158.96, 152.17 – 151.45 (m), 136.65 (t, J = 10.7 Hz), f114.23 (dd, J = 21.1, 2.3 Hz), 112.91 (t, J = 17.5 Hz), 112.40 (dd, J = 21.6, 3.2 Hz), 108.95, 106.89 (dd, J = 27.2, 3.0 Hz), 106.13 (t, J = 16.0 Hz), 92.24 (t, J = 19.3 Hz), 37.92, 23.61, 13.55.

$^{19}$F NMR (376 MHz, CDCl$_3$) (δ): -103.68 (d, J = 8.9 Hz, 2F, Ar-**F**), -108.99 (d, J = 10.4 Hz, 2F, Ar-**F**), -111.66 (d, J = 9.3 Hz, 2F, Ar-**F**).

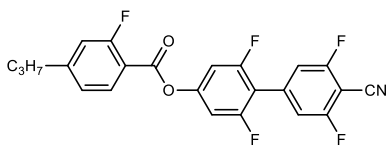

**2** (**2.2.1**): *4'-Cyano-2,3',5',6 tetrafluoro-[1,1' biphenyl]4-yl 2 fluoro-4-propyl benzoate*

Yield: (Shiny white solid) 215 mg, 50%

$R_F$ (DCM): 0.83

¹H NMR (400 MHz, CDCl₃) (δ): 7.99 (t, J = 7.8 Hz, 1H, Ar-**H**), 7.21 (ddd, J = 8.1, 1.8, 1.2 Hz, 2H, Ar-**H**), 7.11 (dd, J = 8.1, 1.6 Hz, 1H, Ar-**H**), 7.08 – 7.01 (m, 3H, Ar-**H**)*, 2.68 (t, J = 6.8 Hz, 2H, Ar-C**H₂**-CH₂), 1.70 (h, J = 7.4 Hz, 2H, CH₂-C**H₂**-CH₃), 0.98 (t, J = 7.3 Hz, 3H, CH₂-C**H₃**).

*Overlapping signals

¹³C{¹H} NMR (101 MHz, CDCl₃) (δ): 164.14 (dd, J = 261.0, 5.3 Hz), 162.05 (d, J = 262.2 Hz), 161.70 (d, J = 4.3 Hz), 159.57 (dd, J = 251.9, 8.4 Hz), 152.83 (d, J = 8.7 Hz), 152.32 (t, J = 14.4 Hz), 136.76 (t, J = 10.7 Hz), 132.44, 124.65 (d, J = 3.3 Hz), 117.16 (d, J = 21.7 Hz), 114.27 (dd, J = 21.0, 2.8 Hz), 108.98, 106.97 (dd, J = 27.0, 3.1 Hz), 92.18 (t, J = 19.3 Hz), 37.89, 23.85, 13.65.

¹⁹F NMR (376 MHz, CDCl₃) (δ): -103.73 (d, J = 9.0 Hz, 2F, Ar-**F**), -107.90 (dd, J = 11.8, 7.5 Hz, 1F, Ar-**F**), -111.94 (d, J = 9.4 Hz, 2F, Ar-**F**).

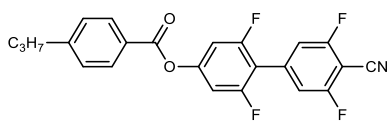

**3** (**2.2.0**): *4'-Cyano-2,3',5',6 tetrafluoro-[1,1' biphenyl]4-yl 4-propyl benzoate*

Yield: (white solid) 281 mg, 68%

R_F (DCM): 0.87

¹H NMR (400 MHz, CDCl₃) (δ): 8.09 (ddd, J = 8.4, 1.8, 1.8 Hz, 2H, Ar-**H**), 7.34 (ddd, J = 8.4, 1.9, 1.9 Hz, 2H, Ar-**H**), 7.22 (ddd, J = 8.2, 1.6, 1.6 Hz, 2H, Ar-**H**), 7.06 – 6.99 (m, 2H, Ar-**H**), 2.70 (t, J = 7.3 Hz, 2H, Ar-C**H₂**-CH₂), 1.70 (h, J = 7.5 Hz, 2H, CH₂-C**H₂**-CH₃), 0.97 (t, J = 7.4 Hz, 3H, CH₂-C**H₃**).

¹³C{¹H} NMR (101 MHz, CDCl₃) (δ): 164.24 – 164.07 (m), 161.53 (d, J = 5.1 Hz), 159.60 (dd, J = 252.0, 8.4 Hz), 152.82 (t, J = 14.5 Hz), 150.13, 136.86 (t, J = 10.5 Hz), 130.44, 128.99, 125.72, 114.27 (dd, J = 21.0, 3.0 Hz), 108.99, 107.25 – 106.73 (m), 92.17 (t, J = 19.3 Hz), 38.16, 24.22, 13.74.

¹⁹F NMR (376 MHz, CDCl₃) (δ): -103.72 (d, J = 8.9 Hz, 2F, Ar-**F**), -112.01 (d, J = 9.5 Hz, 2F, Ar-**F**).

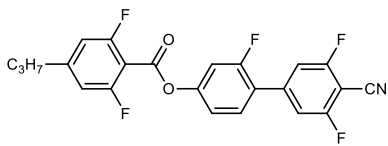

**4** (**2.1.2**): 4'-Cyano-2,3',6 trifluoro-[1,1' biphenyl]4-yl 2,6 difluoro-4-propyl benzoate

Yield: (white powder) 319 mg, 74%

$R_F$ (DCM): 0.83

$^1$H NMR (400 MHz, CDCl$_3$) (δ): 7.41 (t, *J* = 8.6 Hz, 1H, Ar-**H**), 7.26 – 7.08 (m, 5H, Ar-H)*, 6.80 (d$_{apparent}$, *J* = 10.1 Hz, 2H, Ar-**H**), 2.58 (t, *J* = 7.6 Hz, 2H, Ar-C**H$_2$**-CH$_2$), 1.61 (h, *J* = 7.4 Hz, 2H, CH$_2$-C**H$_2$**-CH$_3$), 0.90 (t, *J* = 7.3 Hz, 3H, CH$_2$-C**H$_3$**).

*Overlapping signals

$^{13}$C{$^1$H} NMR (101 MHz, CDCl$_3$) (δ): 164.37 (dd, J = 185.8, 5.2 Hz), 161.80, 161.44 (dd, J = 183.0, 5.5 Hz), 158.81 (d, J = 115.0 Hz), 152.21 (d, J = 11.1 Hz), 151.33 (t, J = 9.9 Hz), 143.04 (t, J = 9.7 Hz), 130.59 (d, J = 3.7 Hz), 123.19 (d, J = 12.8 Hz), 118.61 (d, J = 3.7 Hz), 112.75 (t, J = 3.9 Hz), 112.61 – 112.36 (m)*, 112.23 (d, J = 3.0 Hz) 111.02 (d, J = 25.9 Hz), 109.11, 106.53 (t, J = 16.3 Hz), 91.49 (t, J = 19.1 Hz), 37.90, 23.63, 13.56.

*Overlapping signals

$^{19}$F NMR (376 MHz, CDCl$_3$) (δ): -103.50 (d, J = 9.3 Hz, 2F, Ar-**F**), -109.27 (d, J = 10.4 Hz, 2F, Ar-**F**), -113.29 (t$_{apparent}$, J = 9.9 Hz, 1F, Ar-**F**).

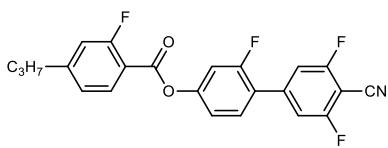

**5** (**2.1.1**): 4'-Cyano-2,3',6 trifluoro-[1,1' biphenyl]4-yl 2 fluoro-4-propyl benzoate

Yield: (shiny white solid) 272 mg, 66%

$R_F$ (DCM): 0.82

$^1$H NMR (400 MHz, CDCl$_3$) (δ): 8.00 (t, J = 7.8 Hz, 1H, Ar-**H**), 7.48 (t, J = 8.6 Hz, 1H, Ar-**H**), 7.32 – 7.27 (m, 1H, Ar-**H**)$^+$, 7.25 – 7.17 (m, 2H, Ar-**H**), 7.11 (dd, J = 8.1, 1.6 Hz, 1H, Ar-**H**), 7.05 (dd, J = 11.9, 1.6 Hz, 1H, Ar-**H**), 2.68 (t, J = 7.6 Hz, 2H, Ar-C**H$_2$**-CH$_2$), 1.70 (h, J = 7.4 Hz, 2H, CH$_2$-C**H$_2$**-CH$_3$), 0.98 (t, J = 7.3 Hz, 3H, CH$_2$-C**H$_3$**).

$^+$Overlapping with solvent peak.

$^{13}$C{$^1$H} NMR (101 MHz, CDCl$_3$) (δ): 162.68 (dd, J = 259.0, 25.7 Hz), 161.80 (d, J = 4.1 Hz), 159.67 (dd, J = 251.2, 8.5 Hz), 152.74 (d, J = 8.6 Hz), 151.87 (t, J = 14.3 Hz), 136.02 (d, J = 8.6 Hz), 133.24, 132.43, 126.88 (d, J = 2.7 Hz), 124.62 (d, J = 3.2 Hz), 118.49 (d, J = 20.9 Hz), 117.15 (d, J = 21.7 Hz), 114.14 (d, J = 9.3 Hz), 113.72, 113.44, 106.82 (dd, J = 27.5, 2.6 Hz), 101.24 (d, J = 15.4 Hz), 77.36, 77.24, 77.04, 76.72, 37.88, 23.86, 13.65.

$^{19}$F NMR (376 MHz, CDCl$_3$) (δ): -103.52 (d, J = 9.3 Hz, 2F, Ar-**F**), -108.16 (dd, J = 12.0, 7.5 Hz, 1F, Ar-**F**), -113.48 (t, J = 9.9 Hz, 1F, Ar-**F**).

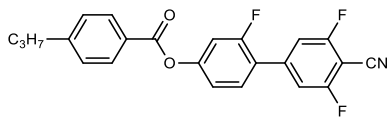

**6** (**2.1.0**): *4'-Cyano-2,3',6 trifluoro-[1,1' biphenyl]4-yl 4-propyl benzoate*

Yield: (white solid) 296 mg, 75%

R$_F$ (DCM): 0.90

$^1$H NMR (400 MHz, CDCl$_3$) (δ): (d $_{(apparent)}$, J = 8.1 Hz, 2H, Ar-**H**), 7.48 (t, J = 8.6 Hz, 1H, Ar-**H**), 7.34 (d $_{(apparent)}$, J = 8.0 Hz, 2H, Ar-**H**), 7.30 – 7.24 (m, 2H, Ar-**H**)*$^+$, 7.23 – 7.14 (m, 1H, Ar-**H**), 2.70 (t, J = 7.6 Hz, 2H, Ar-C**H$_2$**-CH$_2$), 1.70 (h, J = 7.4 Hz, 2H, CH$_2$-C**H$_2$**-CH$_3$), 0.97 (t, J = 7.3 Hz, 3H, CH$_2$-C**H$_3$**).

*Overlapping signals, $^+$overlapping solvent peak.

$^{13}$C{$^1$H} NMR (101 MHz, CDCl$_3$) (δ): 164.59, 163.11 (dd, J = 260.5, 4.9 Hz), 161.81, 158.29, 153.04 (d, J = 11.2 Hz), 149.87, 143.19 (t, J = 9.9 Hz), 130.50 (d, J = 3.6 Hz), 130.39, 128.92, 126.11, 118.76 (d, J = 3.7 Hz), 112.60 (dt, J = 20.9, 3.7 Hz), 111.14 (d, J = 25.5 Hz), 109.15, 38.16, 24.24, 13.75.

$^{19}$F NMR (376 MHz, CDCl$_3$) (δ): -103.55 (d, J = 9.3 Hz, 2F, Ar-**F**), -113.55 (t, J = 9.9 Hz, 1F, Ar-**F**).

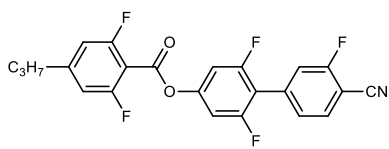

**7** (**1.2.2**): *4'-Cyano-2,3',5' trifluoro-[1,1' biphenyl]4-yl 2,6 difluoro-4-propyl benzoate*

Yield: (shiny white solid) 258 mg, 60%

R$_F$ (DCM): 0.82

<sup>1</sup>H NMR (400 MHz, CDCl<sub>3</sub>) (δ): 7.66 (t, J = 6.9 Hz, 1H, Ar-**H**), 7.36 – 7.27 (m, 2H, Ar-**H**)*, 7.01 – 6.93 (m, 2H, Ar-**H**), 6.84 – 6.75 (m, 2H, Ar-**H**), 2.58 (t, J = 7.4 Hz, 2H, Ar-C**H₂**-CH₂), 1.61 (h, J = 7.5 Hz, 2H, CH₂-C**H₂**-CH₃), 0.90 (t, J = 7.3 Hz, 3H, CH₂-C**H₃**).

*overlapping signals

<sup>13</sup>C{<sup>1</sup>H} NMR (101 MHz, CDCl<sub>3</sub>) (δ): 164.11 (d, J = 260.0 Hz), 162.57 (dd, J = 157.7, 5.8 Hz), 161.13 (d, J = 247.9 Hz), 159.18 (dd, J = 150.5, 5.7 Hz), 151.83 – 151.23 (m)*, 135.91 (d, J = 8.6 Hz), 133.26, 126.87, 118.50 (d, J = 21.0 Hz), 113.71, 112.38 (dd, J = 21.6, 3.3 Hz), 106.74 (dd, J = 27.7, 2.7 Hz), 101.32 (d, J = 15.4 Hz), 37.92, 23.62, 13.56.

*overlapping signals

<sup>19</sup>F NMR (376 MHz, CDCl<sub>3</sub>) (δ): -106.10 (t, J = 6.6 Hz, 1F, Ar-**F**), -109.07 (d, J = 10.4 Hz, 2F, Ar-**F**), -111.96 (d, J = 9.0 Hz, 2F, Ar-**F**).

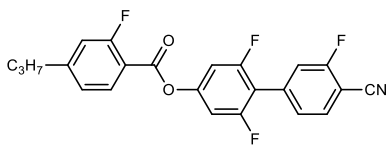

**8 (1.2.1):** 4'-Cyano-2,3',5' trifluoro-[1,1' biphenyl]4-yl 2 fluoro-4-propyl benzoate

Yield: (white solid) 318 mg, 77%

R<sub>F</sub> (DCM): 0.82

<sup>1</sup>H NMR (400 MHz, CDCl<sub>3</sub>) (δ): 7.99 (t, J = 7.8 Hz, 1H, Ar-**H**), 7.72 (t, J = 6.6 Hz, 1H, Ar-**H**), 7.44 – 7.34 (m, 2H, Ar-**H**), 7.11 (dd, J = 8.1, 1.6 Hz, 1H, Ar-**H**), 7.08 – 6.98 (m, 3H, Ar-**H**)*, 2.68 (t, J = 6.8 Hz, 2H, Ar-C**H₂**-CH₂), 1.70 (h, J = 7.3 Hz, 2H, CH₂-C**H₂**-CH₃), 0.97 (t, J = 7.3 Hz, 3H, C**H₂**-CH₃).

*Overlapping signals

<sup>13</sup>C{<sup>1</sup>H} NMR (101 MHz, CDCl<sub>3</sub>) (δ): 162.68 (dd, J = 259.0, 25.7 Hz), 161.80 (d, J = 4.1 Hz), 159.67 (dd, J = 251.2, 8.5 Hz), 152.74 (d, J = 8.6 Hz), 151.87 (t, J = 14.3 Hz), 136.02 (d, J = 8.6 Hz), 133.24, 132.43, 126.88 (d, J = 2.7 Hz), 124.62 (d, J = 3.2 Hz), 118.49 (d, J = 20.9 Hz), 117.15 (d, J = 21.7 Hz), 114.14 (d, J = 9.3 Hz), 106.82 (dd, J = 27.5, 2.6 Hz), 101.24 (d, J = 15.4 Hz), 37.88, 23.86, 13.65.

<sup>19</sup>F NMR (376 MHz, CDCl<sub>3</sub>) (δ): -106.17 (t, J = 6.8 Hz, 1F, Ar-**H**), -107.97 (dd, J = 11.9, 7.6 Hz, 1F, Ar-**H**), -112.25 (d, J = 9.2 Hz, 2F, Ar-**H**).

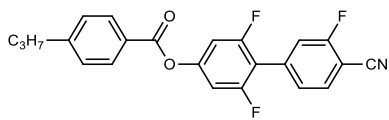

**9 (1.2.0)**: 4'-Cyano-2,3',5' trifluoro-[1,1' biphenyl]4-yl 4-propyl benzoate

Yield: (white solid) 277 mg, 70%

R_F (DCM): 0.85

¹H NMR (400 MHz, CDCl₃) (δ): 8.09 (ddd, J = 8.4, 2.0, 1.8 Hz, 2H, Ar-**H**), 7.72 (t, J = 7.0 Hz, 1H, Ar-**H**), 7.45 – 7.36 (m, 2H, Ar-**H**), 7.35 (ddd, J = 8.6, 1.8, 1.6 Hz, 2H, Ar-**H**), 7.06 – 6.97 (m, 2H, Ar-**H**), 2.70 (t, J = 7.4 Hz, 2H, Ar-C**H₂**-CH₃), 1.70 (h, J = 7.4 Hz, 2H, CH₂-C**H₂**-CH₃), 0.97 (t, J = 7.3 Hz, 3H, CH₂-C**H₃**).

¹³C{¹H} NMR (101 MHz, CDCl₃) (δ): 164.21 (d, J = 18.0 Hz), 161.54, 160.90 (dd, J = 242.2, 8.4 Hz), 152.29 (t, J = 14.2 Hz), 150.05, 136.12 (d, J = 8.9 Hz), 133.24, 130.43, 128.97, 126.88, 125.83, 118.50 (d, J = 21.7 Hz), 113.74, 106.85 (dd, J = 28.2, 2.4 Hz), 101.31, 38.16, 24.22, 13.74.

¹⁹F NMR (376 MHz, CDCl₃) (δ): -106.15 (t, J = 8.0 Hz, 1F, Ar-**F**), -112.33 (d, J = 9.1 Hz, 2F, Ar-**F**).

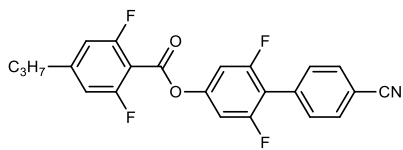

**10 (0.2.2)**: *4'-Cyano-3',5' difluoro-[1,1' biphenyl]4-yl 2,6 difluoro-4-propyl benzoate*

Yield: (fluffy white solid) 335 mg, 80%

R_F (DCM): 0.79

¹H NMR (400 MHz, CDCl₃) (δ): 7.76 (ddd, J = 8.2, 1.7, 1.5 Hz, 2H, Ar-**H**), 7.60 (ddd, J = 8.4, 1.5, 1.5 Hz, 3H, Ar-**H**), 7.07 – 6.98 (m, 2H, Ar-**H**), 6.87 (d (apparent), J = 9.9 Hz, 2H, Ar-**H**), 2.65 (t, J = 7.4 Hz, 2H, Ar-C**H₂**-CH₂), 1.69 (h, J = 7.6 Hz, 2H, CH₂-C**H₂**-CH₃), 0.97 (t, J = 7.3 Hz, 3H, CH₂-C**H₃**).

¹³C{¹H} NMR (101 MHz, CDCl₃) (δ): 161.78 (dd, J = 154.0, 7.4 Hz), 159.23 (dd, J = 137.9, 6.1 Hz), 159.16, 158.57, 158.48, 151.45 (t, J = 9.7 Hz), 150.91 (t, J = 14.4 Hz), 133.42, 132.12, 131.11, 118.57, 114.82 (t, J = 18.4 Hz), 112.47 (d, J = 3.2 Hz), 112.25 (d, J = 3.2 Hz), 106.98 – 106.24 (m)*, 37.91, 23.62, 13.56.

*overlapping signals.

$^{19}$F NMR (376 MHz, CDCl$_3$) (δ): -109.14 (d, J = 10.4 Hz, 2F, Ar-**F**), -112.28 (d, J = 8.8 Hz, 2F, Ar-**F**).

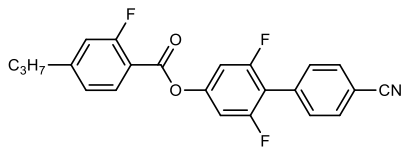

**11** (**0.2.1**): 4'-Cyano-3',5' difluoro-[1,1' biphenyl]4-yl 2 fluoro-4-propyl benzoate

Yield: (white powder) 213 mg, 54%

R$_F$ (DCM): 0.84

$^1$H NMR (400 MHz, CDCl$_3$) (δ): 7.99 (t, J = 7.8 Hz, 1H, Ar-**H**), 7.76 (ddd, J = 8.6, 1.9, 1.6 Hz, 2H, Ar-**H**), 7.60 (ddd, J = 8.6, 1.4, 1.4 Hz, 2H, Ar-**H**), 7.11 (dd, J = 8.0, 1.6 Hz, 1H, Ar-**H**), 7.04 (dd, J = 12.0, 1.4 Hz, 1H, Ar-**H**), 7.02 – 6.97 (m, 2H, Ar-**H**), 2.68 (t, J = 6.8 Hz, 2H, Ar-C**H$_2$**-CH$_2$), 1.70 (h, J = 7.3 Hz, 2H, CH$_2$-C**H$_2$**-CH$_3$), 0.98 (t, J = 7.4 Hz, 3H, CH$_2$-C**H$_3$**).

$^{13}$C{$^1$H} NMR (101 MHz, CDCl$_3$) (δ): 163.84 (d, J = 261.7 Hz), 161.92 (d, J = 3.8 Hz), 160.97, 159.77 (dd, J = 250.5, 8.8 Hz), 152.63 (d, J = 8.7 Hz), 151.39 (t, J = 14.3 Hz), 133.52, 132.44, 132.11, 131.12, 124.60 (d, J = 3.3 Hz), 118.59, 117.25, 117.03, 114.25 (d, J = 9.2 Hz), 112.21, 106.94 – 106.44 (m)., 37.88, 23.86, 13.66.

$^{19}$F NMR (376 MHz, CDCl$_3$) (δ): -108.03 (dd, J = 12.0, 7.5 Hz, 1F, Ar-**F**), -112.56 (d, J = 8.8 Hz, 2F, Ar-**F**).

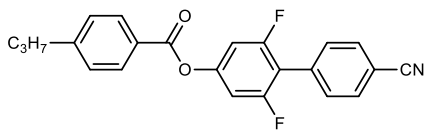

**12** (**0.2.0**): *4'-Cyano-3',5' difluoro-[1,1' biphenyl]4-yl 4-propyl benzoate*

Yield: (fluffy white solid) 230 mg, 61%

R$_F$ (DCM): 0.89

$^1$H NMR (400 MHz, CDCl$_3$) (δ): 8.10 (ddd, J = 8.4, 1.7, 1.6 Hz, 2H, Ar-**H**), 7.76 (ddd, J = 8.7, 1.8, 1.4 Hz, 2H, Ar-**H**), 7.60 (ddd, J = 8.5, 1.4, 1.2 Hz, 2H, Ar-**H**), 7.35 (ddd, J = 8.5, 1.8, 1.6 Hz, 2H, Ar-**H**), 7.04 – 6.93 (m$_{apparent}$, 2H, Ar-**H**), 2.70 (t, J = 7.3 Hz, 2H, Ar-C**H$_2$**-CH$_2$), 1.70 (h, J = 7.5 Hz, 2H, CH$_2$-C**H$_2$**-CH$_3$), 0.97 (t, J = 7.3 Hz, 3H, CH$_2$-C**H$_3$**).

$^{13}$C{$^1$H} NMR (101 MHz, CDCl$_3$) (δ): 164.37, 159.79 (dd, J = 250.3, 8.9 Hz), 151.83 (t, J = 14.3 Hz), 149.97, 133.57, 132.10, 131.12, 130.42, 128.95, 125.94, 118.60, 114.37 (t, J = 18.5 Hz), 112.19, 107.04 – 106.42 (m$_{apparent}$), 38.16, 24.23, 13.75.

$^{19}$F NMR (376 MHz, CDCl$_3$) (δ): -112.65 (d, J = 8.8 Hz, 2F, Ar-**F**).

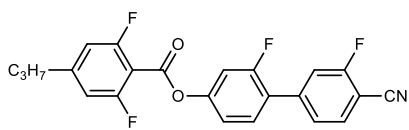

**13** (**1.1.2**): 4'-Cyano-2,3' difluoro-[1,1' biphenyl]4-yl 2,6 difluoro-4-propyl benzoate

Yield: (white solid) 198 mg, 48%

R$_F$ (DCM): 0.83

$^1$H NMR (400 MHz, CDCl$_3$) (δ): 7.71 (t, J = 7.1 Hz, 1H, Ar-**H**), 7.53 – 7.40 (m$_{apparent}$, 3H, Ar-**H**)*, 7.24 – 7.16 (m$_{apparent}$, 2H)*, 6.91 – 6.84 (m$_{apparent}$, 2H), 2.65 (t, J = 7.4 Hz, 2H, Ar-C**H$_2$**-CH$_2$), 1.68 (h, J = 7.4 Hz, 2H, CH$_2$-C**H$_2$**-CH$_3$), 0.97 (t, J = 7.3 Hz, 3H, CH$_2$-C**H$_3$**).

*overlapping signals.

$^{13}$C{$^1$H} NMR (101 MHz, CDCl$_3$) (δ): 164.35, 161.77 (dd, J = 258.6, 6.1 Hz), 161.7, 159.47 (d, J = 252.6 Hz), 159.4, 151.75 (d, J = 11.1 Hz), 151.22 (t, J = 9.9 Hz), 142.29 (d, J = 8.4 Hz), 130.77 (d, J = 4.0 Hz), 125.33 (t, J = 3.3 Hz), 124.08 (d, J = 12.8 Hz), 118.42 (d, J = 3.7 Hz), 116.92 (dd, J = 20.7, 3.9 Hz), 113.85, 112.32 (dd, J = 21.7, 3.2 Hz), 110.85 (d, J = 25.8 Hz), 106.64, 100.60 (d, J = 15.6 Hz), 37.89, 23.63, 13.56.

$^{19}$F NMR (376 MHz, CDCl$_3$) (δ): -106.08 (t$_{apparent}$, J = 7.9 Hz, 1F, Ar-**F**), -109.34 (d, J = 10.3 Hz, 2F, Ar-**F**), -113.76 (t$_{apparent}$, J = 9.8 Hz, 1F, Ar-**F**).

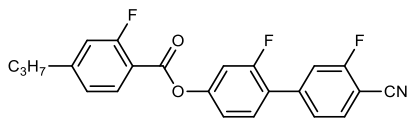

**14** (**1.1.1**): *4'-Cyano-2,3' difluoro- [1,1' biphenyl]4-yl 2 fluoro-4-propyl benzoate*

Yield: (shiny white solid) 245 mg, 62%

R$_F$ (DCM): 0.85

$^1$H NMR (400 MHz, CDCl$_3$) (δ): 7.94 (t, J = 8.3 Hz, 1H, Ar-**H**), 7.64 (t, J = 7.2 Hz, 1H, Ar-**H**), 7.39 (m, 3H, Ar-**H**)*, 7.12 (m, 2H, Ar-**H**)*, 7.04 (d$_{apparent}$, J = 7.9 Hz, 1H, Ar-**H**), 6.98 (d$_{apparent}$, J = 11.8 Hz, 1H, Ar-**H**), 2.61 (t, J = 7.6 Hz, 2H, Ar-C**H$_2$**-CH$_2$), 1.63 (h, J = 7.4 Hz, 2H, CH$_2$-C**H$_2$**-CH$_3$), 0.91 (t, J = 7.3 Hz, 3H, CH$_2$-C**H$_3$**).

$^{19}$F NMR (376 MHz, CDCl$_3$) (δ): -104.73 (t, J = 9.4 Hz, 1F, Ar-**F**), -108.23 (t, J = 10.1 Hz, 1F, Ar-**F**), -113.96 (t, J = 10.3 Hz, 1F, Ar-**F**).

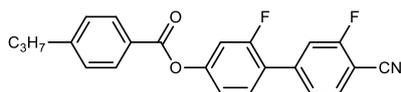

**15 (1.1.0)**: 4'-Cyano-2,3' difluoro-[1,1' biphenyl]4-yl-4-propyl benzoate

Yield: (white solid) 204 mg, 54 %

$R_F$ (DCM): 0.91

$^1$H NMR (400 MHz, CDCl$_3$) (δ): 8.11 (ddd, J = 8.3, 1.9, 1.7 Hz, 2H, Ar-**H**), 7.71 (t, J = 6.9 Hz, 1H, Ar-**H**), 7.52 – 7.41 (m, 3H, Ar-**H**)*, 7.34 (ddd, J = 8.4, 1.6, 1.5 Hz, 2H, Ar-**H**), 7.20 – 7.13 (m, 2H, Ar-**H**)*, 2.70 (t, J = 6.8 Hz, 2H, Ar-C**H$_2$**-CH$_2$), 1.70 (h, J = 7.4 Hz, 2H, CH$_2$-C**H$_2$**-CH$_3$), 0.97 (t, J = 7.3 Hz, 3H, CH$_2$-CH$_3$).

*overlapping signals.

$^{13}$C{$^1$H} NMR (101 MHz, CDCl$_3$) (δ): 164.68, 163.08 (d, J = 259.0 Hz), 159.56 (d, J = 251.9 Hz), , 152.56 (d, J = 11.2 Hz), 149.78, 142.45 (d, J = 8.4 Hz), 133.47, 130.68 (d, J = 4.0 Hz), 130.38, 128.90, 126.21, 125.31 (t$_{apparent}$, J = 3.5 Hz), 123.64 (d, J = 12.8 Hz), 118.58 (d, J = 3.7 Hz), 116.99 (d, J = 3.4 Hz), 116.80 (d, J = 3.5 Hz), 113.89, 111.10, 110.85, 100.52 (d, J = 15.7 Hz), 38.16, 24.24, 13.75.

$^{19}$F NMR (376 MHz, CDCl$_3$) (δ): -106.09 (t$_{apparent}$, J = 8.1 Hz, 1F, Ar-**F**), -114.02 (t$_{apparent}$, J = 9.7 Hz, 1F, Ar-**F**).

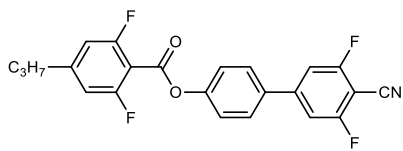

**16 (2.0.2)**: *4'-Cyano-3',5' difluoro-[1,1' biphenyl]4-yl 2,6 difluoro-4-propyl benzoate*

Yield: (fluffy white solid) 314 mg, 76%

$R_F$ (DCM): 0.82

$^1$H NMR (400 MHz, CDCl$_3$) (δ): 7.63 (d$_{apparent}$, J = 8.2 Hz, 2H, Ar-**H**), 7.40 (d$_{apparent}$, J = 8.4 Hz, 2H, Ar-**H**), 7.32 – 7.23 (m$_{apparent}$, 2H, Ar-**H**)+, 6.87 (d$_{apparent}$, J = 10.1 Hz, 2H, Ar-**H**), 2.65 (t, J = 7.6 Hz, 2H, Ar-C**H$_2$**-CH$_2$), 1.69 (h, J = 7.4 Hz, 2H, CH$_2$-C**H$_2$**-CH$_3$), 0.98 (t, J = 7.3 Hz, 3H, CH$_2$-C**H$_3$**).

+overlapping with solvent peak

$^{13}$C{$^1$H} NMR (101 MHz, CDCl$_3$) (δ): 164.71 (dd, J = 261.5, 5.3 Hz), 162.61 – 159.66 (m)*, 151.77, 150.95 (t$_{apparent}$, J = 9.8 Hz), 148.40 (t$_{apparent}$, J = 10.6 Hz), 135.13, 128.40, 122.75, 112.27 (dd, J = 24.1, 2.6 Hz), 110.61 (dd, J = 20.1, 3.4 Hz), 109.27, 106.96 (t$_{apparent}$, J = 16.5 Hz), 90.90 (t$_{apparent}$, J = 19.9 Hz), 37.88, 23.65, 13.56.

*overlapping signals.

$^{19}$F NMR (376 MHz, CDCl$_3$) (δ): -103.23 (d$_{apparent}$, J = 9.5 Hz, 2F, Ar-**F**), -109.57 (d$_{apparent}$, J = 10.2 Hz, 2F, Ar-**F**).

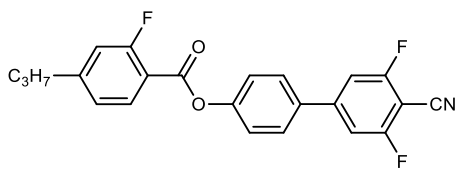

**17 (2.0.1)**: *4'-Cyano-3',5' difluoro-[1,1' biphenyl]4-yl 2 fluoro-4-propyl benzoate*

Yield: (white solid) 225 mg, 57%

R$_F$ (DCM): 0.83

$^1$H NMR (400 MHz, CDCl$_3$) (δ): 7.93 (t, J = 7.8 Hz, 1H, Ar-**H**), 7.54 (ddd, J = 8.8, 2.7, 2.1 Hz, 2H, Ar-**H**), 7.29 (ddd, J = 8.8, 3.0, 2.3 Hz, 2H, Ar-**H**), 7.19 (ddd, J = 8.0, 1.8, 1.8 Hz, 2H, Ar-**H**), 7.02 (dd, J = 8.1, 1.6 Hz, 1H, Ar-**H**), 6.96 (dd, J = 11.9, 1.6 Hz, 1H, Ar-**H**), 2.60 (t, J = 6.8 Hz, 2H, Ar-C**H$_2$**-CH$_2$), 1.62 (h, J = 7.3 Hz, 2H, CH$_2$-C**H$_2$**-CH$_3$), 0.89 (t, J = 7.3 Hz, 3H, CH$_2$-C**H$_3$**).

$^{13}$C{$^1$H} NMR (101 MHz, CDCl$_3$) (δ): 164.71 (dd, J = 221.0, 5.2 Hz), 162.50 (d, J = 261.2 Hz), 162.12 (d, J = 5.2 Hz), 152.23 (d, J = 8.6 Hz), 152.11, 148.49 (t$_{apparent}$, J = 9.8 Hz), 134.79 (t$_{apparent}$, J = 2.8 Hz), 132.41, 128.33, 124.53, 124.50, 122.85, 124.51 (d, J = 3.3 Hz), 117.06 (d, J = 21.8 Hz), 114.78 (d, J = 9.5 Hz), 110.53 (dd, J = 20.5, 3.3 Hz), 109.30, 90.78 (t$_{apparent}$, J = 19.4 Hz), 37.85, 23.89, 13.67.

$^{19}$F NMR (376 MHz, CDCl$_3$) (δ): -103.33 (d, J = 9.4 Hz, 2F, Ar-**F**), -108.44 (dd, J = 12.0, 7.7 Hz, 1F, Ar-**F**).

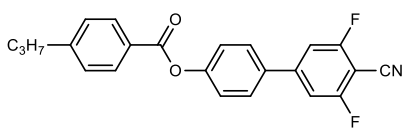

**18 (2.0.0)**: *4'-Cyano-3',5' difluoro-[1,1' biphenyl]4-yl-4-propyl benzoate*

Yield: (white solid) 230 mg, 61%

R$_F$ (DCM): 0.87

$^1$H NMR (400 MHz, CDCl$_3$) (δ): 8.05 (ddd, J = 8.3, 1.9, 1.7 Hz, 2H, Ar-**H**), 7.55 (ddd, J = 8.7, 2.7, 2.3 Hz, 2H, Ar-**H**), 7.31 – 7.23 (m, 4H, Ar-**H**)*⁺, 7.19 (d$_{apparent}$, J = 8.5 Hz, 2H, Ar-**H**), 2.62 (t, J = 6.8 Hz, 2H, Ar-C**H$_2$**-CH$_2$), 1.63 (h, J = 7.4 Hz, 2H, CH$_2$-C**H$_2$**-CH$_3$), 0.90 (t, J = 7.3 Hz, 3H, CH$_2$-C**H$_3$**).

*overlapping peaks, ⁺overlapping with solvent peak

$^{13}$C{$^1$H} NMR (101 MHz, CDCl$_3$) (δ): 165.00, 163.43 (dd, J = 260.8, 5.4 Hz), 152.48, 149.59, 148.54 (t$_{apparent}$, J = 9.8 Hz), 134.67 (t$_{apparent}$, J = 2.0 Hz), 130.34, 128.86, 128.34, 126.51, 122.90, 110.52 (dd, J = 20.3, 3.4 Hz), 109.31, 90.69 (t$_{apparent}$, J = 19.1 Hz), 38.15, 24.26, 13.76.

$^{19}$F NMR (376 MHz, CDCl$_3$) (δ): -103.31 (d$_{apparent}$, J = 9.3 Hz, 2F, Ar-**F**).

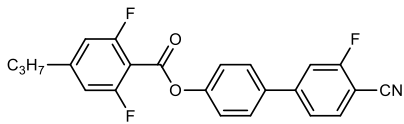

**19** (**1.0.2**): *4'-Cyano-3' fluoro-[1,1' biphenyl]4-yl 2,6 difluoro-4-propyl benzoate*

Yield: (shiny white solid) 280 mg, 71%

R$_F$ (DCM): 0.79

$^1$H NMR (400 MHz, CDCl$_3$) (δ): 7.70 (t, J = 7.8 Hz, 1H, Ar-**H**), 7.64 (ddd, J = 8.6, 2.6, 1.9 Hz, 2H, Ar-**H**), 7.48 (dd, J = 8.1, 1.7 Hz, 1H, Ar-**H**), 7.43 (dd, J = 10.2, 1.3 Hz, 1H, Ar-**H**), 7.39 (ddd, J = 8.7, 2.9, 1.9 Hz, 2H), 6.86 (d$_{apparent}$, J = 9.9 Hz, 2H, Ar-**H**), 2.65 (t, J = 7.6 Hz, 2H, Ar-C**H$_2$**-CH$_2$), 1.68 (h, J = 7.4 Hz, 2H, CH$_2$-C**H$_2$**-CH$_3$), 0.97 (t, J = 7.3 Hz, 3H, CH$_2$-C**H$_3$**).

$^{13}$C{$^1$H} NMR (101 MHz, CDCl$_3$) (δ): 164.76 (d, J = 259.7 Hz), 162.19 (dd, J = 258.3, 5.8 Hz), 150.83, 150.78 (t$_{apparent}$, J = 10.1 Hz), 147.59 (d, J = 7.9 Hz), 136.07 (d, J = 1.5 Hz), 133.84, 128.45, 123.40 (d, J = 3.3 Hz), 122.58, 114.86 (d, J = 20.3 Hz), 114.00, 112.22 (dd, J = 21.1, 2.7 Hz), 100.06 (d, J = 15.8 Hz), 37.88, 23.66, 13.57.

$^{19}$F NMR (376 MHz, CDCl$_3$) (δ): -105.91 (t$_{apparent}$, J = 7.4 Hz, 1F, Ar-**F**), -109.64 (d, J = 10.1 Hz, 2F, Ar-**F**).

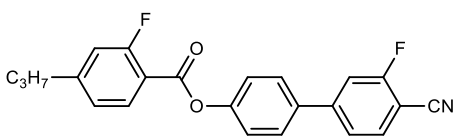

**20** (**1.0.1**): *4'-Cyano-3' fluoro-[1,1' biphenyl]4-yl 2 fluoro-4-propyl benzoate*

Yield: (White solid) 155 mg, 41%

R$_F$ (DCM): 0.84

$^1$H NMR (400 MHz, CDCl$_3$) (δ): 8.02 (t, J = 7.8 Hz, 1H, Ar-**H**), 7.69 (dd, J = 8.1, 6.6 Hz, 1H, Ar-**H**), 7.63 (ddd, J = 8.8, 2.7, 2.1 Hz, 2H, Ar-**H**), 7.48 (dd, J = 8.1, 1.7 Hz, 1H, Ar-**H**), 7.43 (dd, J = 10.1, 1.6 Hz, 1H, Ar-**H**), 7.40 – 7.33 (m, 2H, Ar-**H**), 7.10 (dd, J = 8.0, 1.6 Hz, 1H, Ar-**H**), 7.04 (dd, J = 11.8, 1.6 Hz, 1H, Ar-**H**), 2.68 (t, J = 6.8 Hz, 2H, Ar-C**H$_2$**-CH$_2$), 1.70 (h, J = 7.3 Hz, 2H, CH$_2$-C**H$_2$**-CH$_3$), 0.98 (t, J = 7.3 Hz, 3H, CH$_2$-C**H$_3$**).

$^{13}$C{$^1$H} NMR (101 MHz, CDCl$_3$) (δ): 164.76 (d, J = 258.9 Hz), 162.63 (d, J = 4.1 Hz), 162.39 (d, J = 261.5 Hz), 152.11 (d, J = 8.7 Hz), 151.65, 147.67 (d, J = 8.1 Hz), 135.77 (d, J = 1.7 Hz), 133.82, 132.41, 128.39, 124.48 (d, J = 3.2 Hz), 123.38 (d, J = 3.2 Hz), 122.68, 114.93 (d, J = 9.2 Hz), 114.79 (d, J = 20.5 Hz), 114.93 (d, J = 9.2 Hz), 114.79 (d, J = 20.5 Hz), 99.97 (d, J = 15.6 Hz), 37.85, 23.89, 13.67.

$^{19}$F NMR (376 MHz, CDCl$_3$) (δ): -105.97 (dd, J = 10.1, 6.6 Hz, 1F, Ar-**F**), -108.50 (dd, J = 11.8, 7.5 Hz, 1F, Ar-**F**).

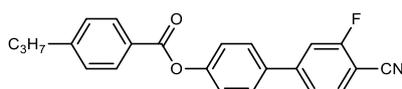

**21** (**1.0.0**): *4'-Cyano-3' fluoro-[1,1' biphenyl]4-yl-4-propyl benzoate*

Yield: (white solid) 233 mg, 65%

R$_F$ (DCM): 0.87

$^1$H NMR (400 MHz, CDCl$_3$) (δ): 8.13 (ddd, J = 8.4, 1.8, 1.6 Hz, 2H, Ar-**H**), 7.69 (t, J = 8.1 Hz, 1H, Ar-**H**), 7.63 (ddd, J = 8.8, 2.6, 2.3 Hz, 2H, Ar-**H**), 7.48 (dd, J = 8.1, 1.7 Hz, 1H, Ar-**H**), 7.43 (dd, J = 10.1, 1.6 Hz, 1H, Ar-**H**), 7.37 – 7.31 (m, 4H, Ar-**H**)*, 2.70 (t, J = 7.6 Hz, 2H, Ar-C**H$_2$**-CH$_2$), 1.71 (h, J = 7.5 Hz, 2H, CH$_2$-C**H$_2$**-CH$_3$), 0.98 (t, J = 7.3 Hz, 3H, CH$_2$-C**H$_3$**).

*overlapping peaks

$^{13}$C{$^1$H} NMR (101 MHz, CDCl$_3$) (δ): 165.08, 163.48 (d, J = 258.7 Hz), 152.02, 149.51, 147.70 (d, J = 8.1 Hz), 135.60 (d, J = 1.9 Hz), 133.82, 130.33, 128.85, 128.39, 123.37 (d, J = 3.2 Hz), 122.73, 114.78 (d, J = 20.3 Hz), 114.05, 100.00, 99.92 (d, J = 15.6 Hz), 38.15, 24.27, 13.77.

$^{19}$F NMR (376 MHz, CDCl$_3$) (δ): -105.99 (t$_{apparent}$, J = 7.6 Hz, 1F, Ar-**F**).

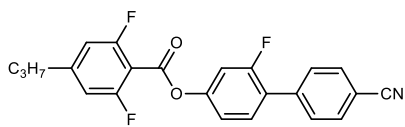

**22** (**0.1.2**): *4'-Cyano-2 fluoro-[1,1' biphenyl]4-yl 2,6 difluoro-4-propyl benzoate*

Yield: (shiny white solid) 256 mg, 65%

R$_F$ (DCM): 0.83

$^1$H NMR (400 MHz, CDCl$_3$) (δ): 7.74 (ddd, J = 8.6, 1.9, 1.6 Hz, 2H, Ar-**H**), 7.68 – 7.63 (m$_{apparent}$, 2H, Ar-**H**), 7.48 (t, J = 8.6 Hz, 1H, Ar-**H**), 7.22 – 7.14 (m, 2H, Ar-**H**)*, 6.87 (d$_{apparent}$, J = 9.4 Hz,

2H, Ar-**H**), 2.65 (t, J = 7.3 Hz, 2H, Ar-C**H₂**-CH₂), 1.68 (h, J = 7.5 Hz, 2H, CH₂-C**H₂**-CH₃), 0.97 (t, J = 7.3 Hz, 3H, CH₂-C**H₃**).

*overlapping peaks

¹³C{¹H} NMR (101 MHz, CDCl₃) (δ): 162.48 (dd, J = 258.1, 6.0 Hz), 159.58, 159.55 (d, J = 251.1 Hz), 151.56 – 150.89 (m), 139.69, 132.34, 130.90 (d, J = 4.1 Hz), 129.67 (d, J = 3.2 Hz), 125.24 (d, J = 13.1 Hz), 118.71, 118.24 (d, J = 3.7 Hz), 112.30 (dd, J = 21.6, 3.1 Hz), 111.63, 110.68 (d, J = 26.0 Hz), 106.76 (t_apparent, J = 16.5 Hz), 37.89, 23.64, 13.57.

¹⁹F NMR (376 MHz, CDCl₃) (δ): -109.42 (d, J = 10.2 Hz, 2F, Ar-**F**), -114.23 (t_apparent, J = 9.8 Hz, 1F, Ar-**F**).

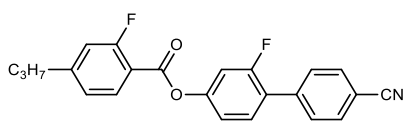

**23** (**0.1.1**): *4'-Cyano-2 fluoro-[1,1' biphenyl]4-yl 2 fluoro-4-propyl benzoate*

Yield: (white solid) 203 mg, 54%

R_F (DCM): 0.84

¹H NMR (400 MHz, CDCl₃) (δ): 8.01 (t, J = 7.8 Hz, 1H, Ar-**H**), 7.75 (ddd, J = 8.6, 1.8, 1.3 Hz, 2H, Ar-**H**), 7.69 – 7.63 (m_apparent, 2H, Ar-**H**), 7.48 (t, J = 8.6 Hz, 1H, Ar-**H**), 7.19 – 7.13 (m, 2H, Ar-**H**)*, 7.10 (dd, J = 8.1, 1.6 Hz, 1H, Ar-**H**), 7.04 (dd, J = 11.9, 1.6 Hz, 1H, Ar-**H**), 2.68 (t, J = 6.8 Hz, 2H, Ar-C**H₂**-CH₂), 1.69 (h, J = 7.3 Hz, 2H, CH₂-C**H₂**-CH₃), 0.97 (t, J = 7.4 Hz, 3H, CH₂-C**H₃**).

*overlapping peaks

¹³C{¹H} NMR (101 MHz, CDCl₃) (δ): 162.44 (d, J = 261.5 Hz), 162.29, 161.21, 159.56 (d, J = 251.3 Hz), 152.32 (d, J = 8.7 Hz), 151.67 (d, J = 11.0 Hz), 139.78, 132.42, 132.33, 130.83 (d, J = 4.1 Hz), 129.66 (d, J = 3.2 Hz), 124.95 (d, J = 13.1 Hz), 124.53 (d, J = 3.2 Hz), 118.74, 118.35 (d, J = 3.7 Hz), 117.09 (d, J = 21.7 Hz), 114.62 (d, J = 9.5 Hz), 111.57, 110.77 (d, J = 25.9 Hz), 37.86, 23.88, 13.66.

¹⁹F NMR (376 MHz, CDCl₃) (δ): -108.29 (dd, J = 12.0, 7.7 Hz, 1F, Ar-**F**), -114.45 (t_apparent, J = 9.7 Hz, 1F, Ar-**F**).

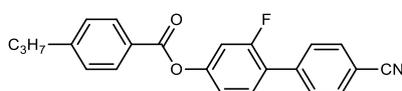

**24** (**0.1.0**): *4'-Cyano-2 fluoro-[1,1' biphenyl]4-yl-4-propyl benzoate*

Yield: (shiny white solid) 220 mg, 61%

$R_F$ (DCM): 0.87

$^1$H NMR (400 MHz, CDCl$_3$) (δ): 8.12 (ddd, J = 8.3, 1.8, 1.8 Hz, 2H, Ar-**H**), 7.74 (ddd, J = 8.5, 1.9, 1.8 Hz, 2H, Ar-**H**), 7.69 – 7.64 (m, 2H, Ar-**H**)*, 7.47 (t, J = 8.3 Hz, 1H, Ar-**H**), 7.34 (ddd, J = 8.3, 1.9, 1.9 Hz, 2H, Ar-**H**), 7.18 – 7.10 (m$_{apparent}$, 1H, Ar-**H**), 2.70 (t, J = 7.5 Hz, 2H, Ar-C**H$_2$**-CH$_2$), 1.70 (h, J = 7.2 Hz, 2H, CH$_2$-C**H$_2$**-CH$_3$), 0.98 (t, J = 7.3 Hz, 3H, CH$_2$-C**H$_3$**).

*overlapping peaks

$^{13}$C{$^1$H} NMR (101 MHz, CDCl$_3$) (δ): 164.77, 159.58 (d, J = 251.1 Hz), 152.08 (d, J = 11.1 Hz), 149.69, 139.82, 132.33, 130.82 (d, J = 4.2 Hz), 130.37, 129.66 (d, J = 3.3 Hz), 128.89, 124.79 (d, J = 13.2 Hz), 124.85, 124.72, 118.75, 118.40 (d, J = 3.7 Hz), 111.54, 110.80 (d, J = 25.7 Hz), 38.15, 24.25, 13.77.

$^{19}$F NMR (376 MHz, CDCl$_3$) (δ): -114.50 (t$_{apparent}$, *J* = 9.8 Hz, 1F, Ar-**F**).

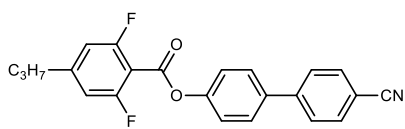

**25** (**0.0.2**): *4'-Cyano-[1,1' biphenyl]4-yl 2,6 difluoro-4-propyl benzoate*

Yield: (white solid) 271 mg, 72%

$R_F$ (DCM): 0.80

$^1$H NMR (400 MHz, CDCl$_3$) (δ): 7.76 – 7.66 (m$_{apparent}$, 2H, Ar-**H**), 7.64 (ddd, J = 8.7, 2.8, 2.1 Hz, 2H, Ar-**H**), 7.37 (ddd, J = 8.6, 2.7, 2.1 Hz, 2H, Ar-**H**), 6.86 (d$_{apparent}$, J = 9.3 Hz, 2H, Ar-**H**), 2.64 (t, J = 7.6 Hz, 2H, Ar-C**H$_2$**-CH$_2$), 1.68 (h, J = 7.6 Hz, 2H, CH$_2$-C**H$_2$**-CH$_3$), 0.97 (t, J = 7.4 Hz, 3H, CH$_2$-C**H$_3$**).

$^{13}$C{$^1$H} NMR (101 MHz, CDCl$_3$) (δ): 160.91 (dd, J = 257.5, 6.9 Hz), 160.01, 150.88, 150.73 (t$_{apparent}$, J = 10.4 Hz), 144.73, 137.26, 132.68, 128.45, 127.75, 122.39, 118.86, 112.23 (dd, J = 21.8, 3.1 Hz), 111.14, 107.20 (t$_{apparent}$, J = 16.9 Hz)., 37.87, 23.66, 13.57.

$^{19}$F NMR (376 MHz, CDCl$_3$) (δ): -109.71 (d$_{apparent}$, J = 10.0 Hz, 2F, Ar-**F**).

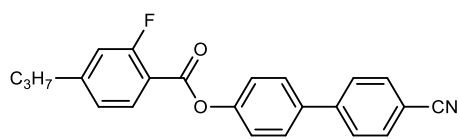

**26** (**0.0.1**): *4'-Cyano-[1,1' biphenyl]4-yl 2 fluoro-4-propyl benzoate*

Yield: (shiny white solid) 248 mg, 69%

R_F (DCM): 0.80

$^1$H NMR (400 MHz, CDCl$_3$) (δ): 8.02 (t, J = 7.8 Hz, 1H, Ar-**H**), 7.77 – 7.66 (m, 4H, Ar-**H**)*, 7.64 (ddd, J = 8.8, 2.8, 2.1 Hz, 2H, Ar-**H**), 7.35 (ddd, J = 8.7, 2.7, 2.1 Hz, 2H, Ar-**H**), 7.10 (dd, J = 8.1, 1.6 Hz, 1H, Ar-**H**), 7.04 (dd, J = 11.8, 1.6 Hz, 1H, Ar-**H**), 2.68 (t, J = 6.8 Hz, 2H, Ar-C**H₂**-CH₂), 1.70 (h, J = 7.5 Hz, 2H, CH₂-C**H₂**-CH₃), 0.98 (t, J = 7.3 Hz, 3H, CH₂-C**H₃**).

*overlapping peaks

$^{13}$C{$^1$H} NMR (101 MHz, CDCl$_3$) (δ): 162.71 (d, J = 4.4 Hz), 162.48 (d, J = 261.4 Hz), 152.01 (d, J = 8.7 Hz), 151.19, 144.82, 136.97, 132.68, 132.41, 128.39, 127.73, 124.46 (d, J = 3.3 Hz), 122.50, 118.89, 117.14 (d, J = 22.1 Hz), 115.01 (d, J = 9.4 Hz), 111.07, 37.85, 23.90, 13.67.

$^{19}$F NMR (376 MHz, CDCl$_3$) (δ): -108.55 (dd, J = 11.8, 7.5 Hz, 1F, Ar-**F**).

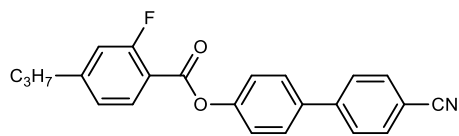

**27** (**0.0.0**) *4'-Cyano-[1,1' biphenyl]4-yl-4-propyl benzoate*

Yield: (shiny white solid) 255 mg, 75%

R_F (DCM): 0.85

$^1$H NMR (400 MHz, CDCl$_3$) (δ): 8.13 (ddd, J = 8.3, 1.7, 1.7 Hz, 2H, Ar-**H**), 7.79 – 7.66 (m, 4H, Ar-**H**)*, 7.64 (ddd, J = 8.5, 2.8, 2.8 Hz, 2H, Ar-**H**), 7.38 – 7.29 (m, 4H, Ar-**H**)*, 2.69 (t, J = 7.7 Hz, 2H, Ar-C**H₂**-CH₂), 1.70 (h, J = 7.2 Hz, 2H, CH₂-C**H₂**-CH₃), 0.97 (t, J = 7.3 Hz, 3H, CH₂-C**H₃**).

*overlapping peaks

$^{13}$C{$^1$H} NMR (101 MHz, CDCl$_3$) (δ): 165.19, 151.54, 149.42, 144.88, 136.83, 132.68, 130.31, 128.82, 128.40, 127.72, 126.72, 122.54, 118.89, 111.03, 38.14, 24.27, 13.76.

### 3.4 Example Structural Characterisation

Below are example $^1$H, $^{13}$C{$^1$H}, and $^{19}$F NMR spectra. Full analysed and raw data for all **1-27** is openly available from the University of Leeds Data Repository at https://doi.org/10.5518/1573.

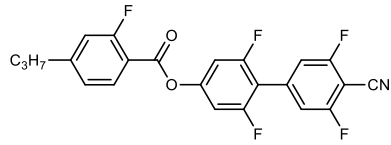

**2 (2.2.1)**

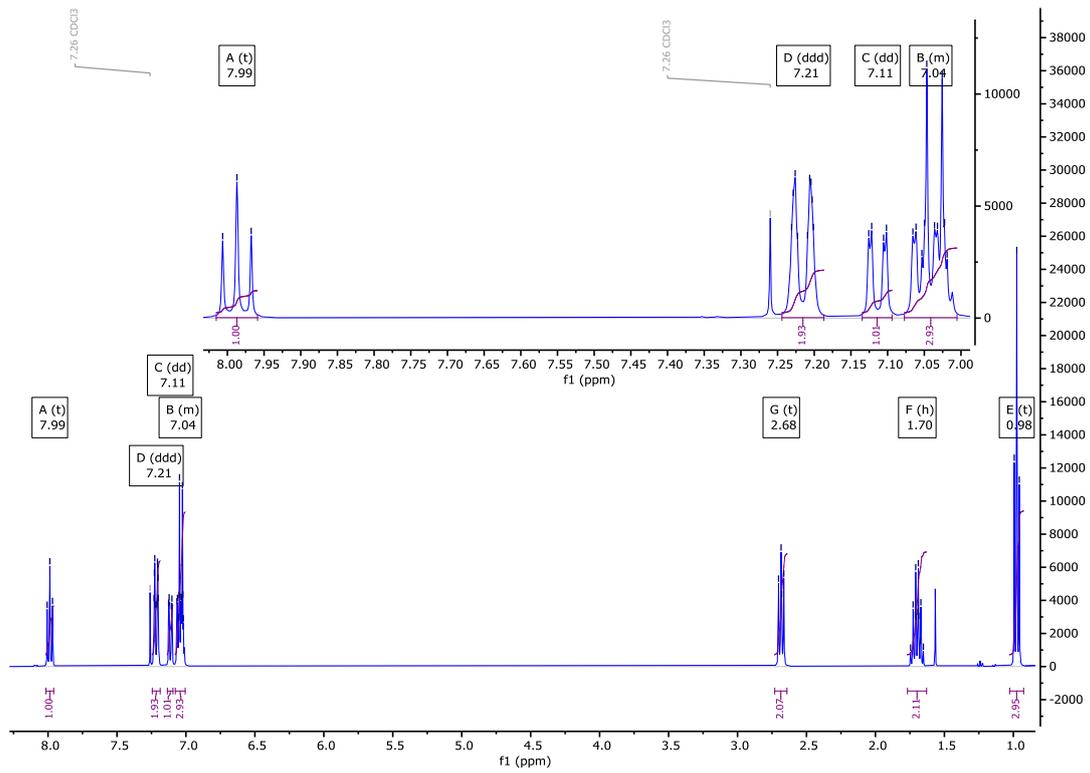

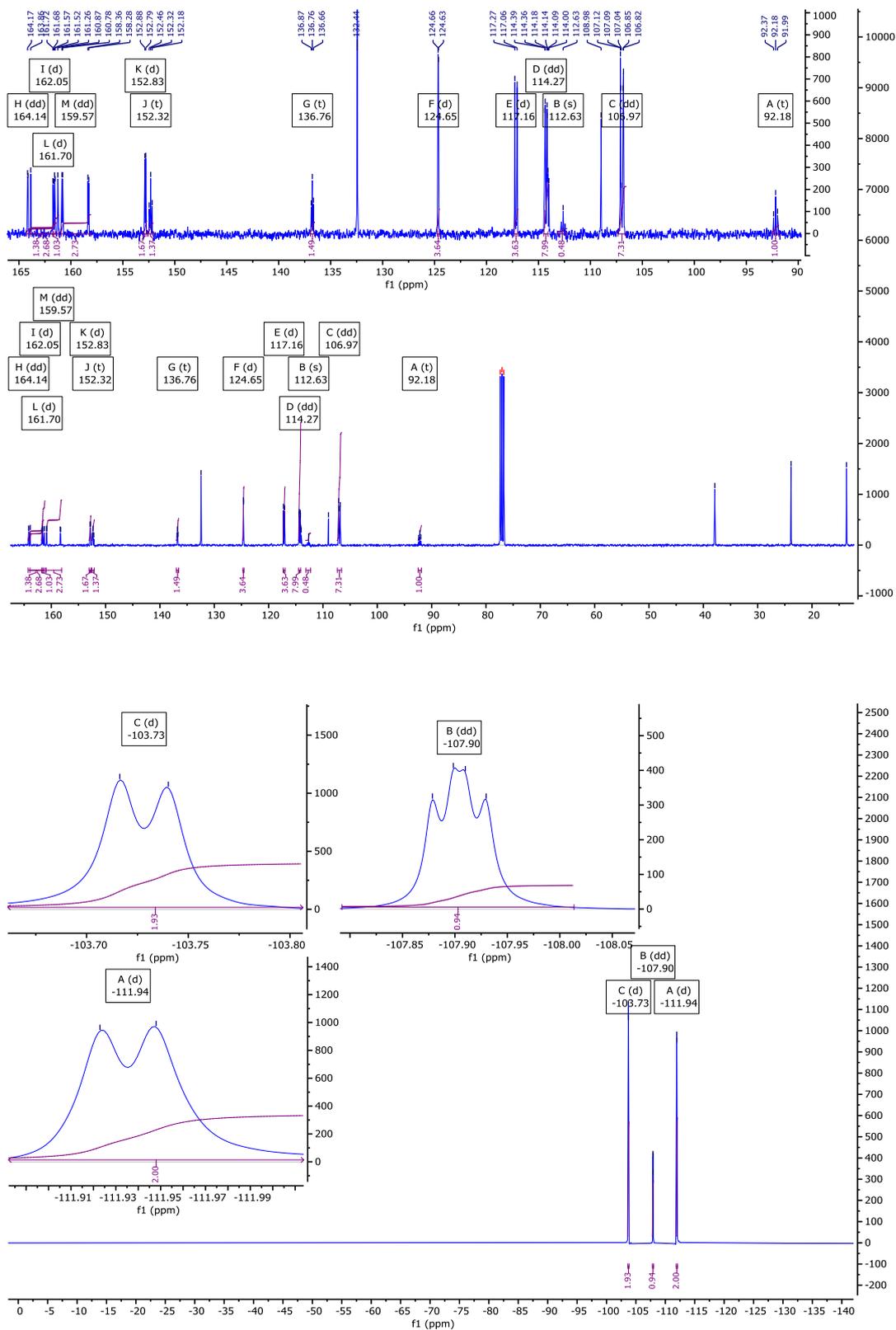

**Figure S6**: Chemical structure, NMR spectra ($^1$H (top), $^{13}$C[1H] (middle), and $^{19}$F (bottom)) spectra for **2 (2.2.1)**.

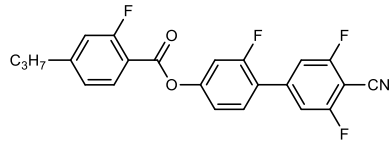

**5 (2.1.1)**

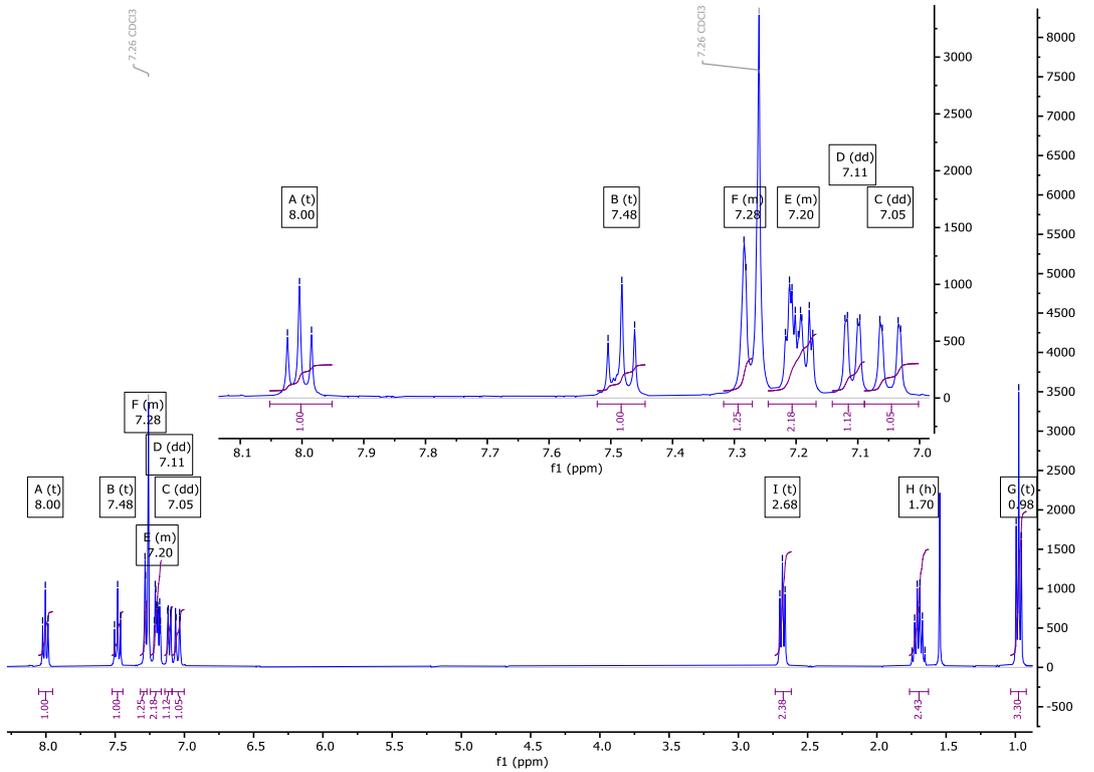
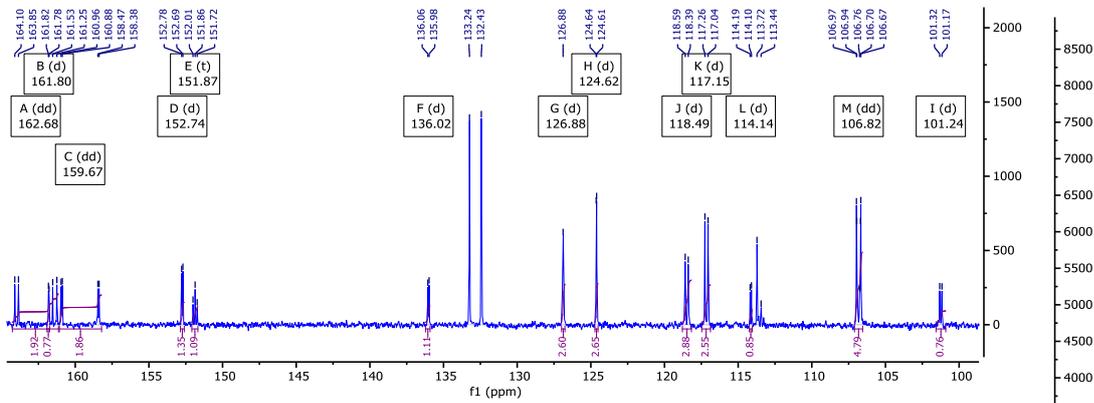
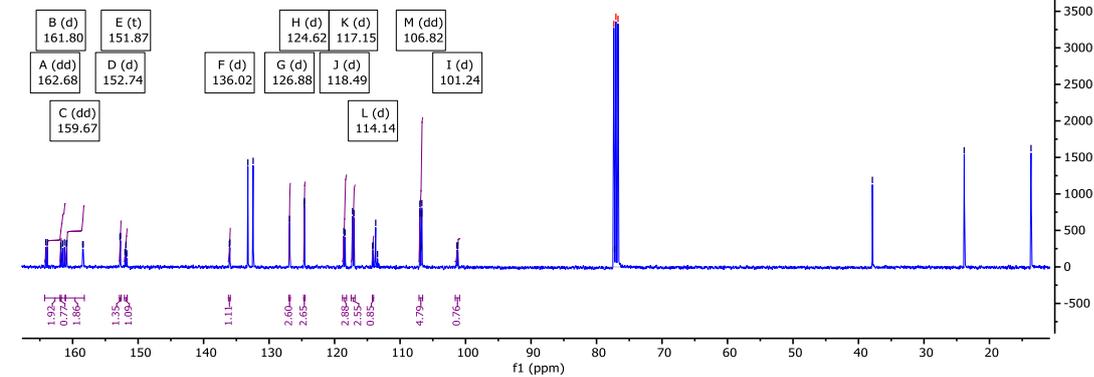

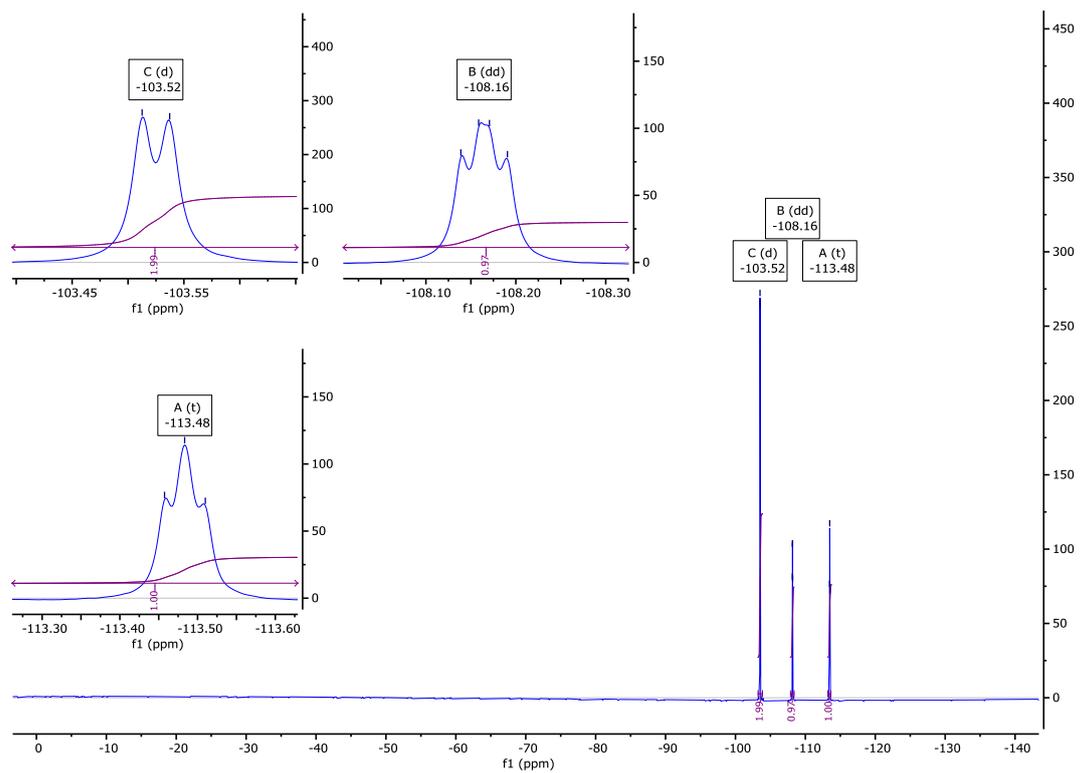

**Figure S7**: Chemical structure, NMR spectra ($^1$H (top), $^{13}$C[1H] (middle), and $^{19}$F (bottom)) spectra for **5 (2.1.1)**.

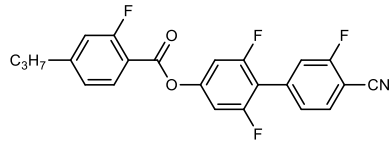

**8 (1.2.1)**

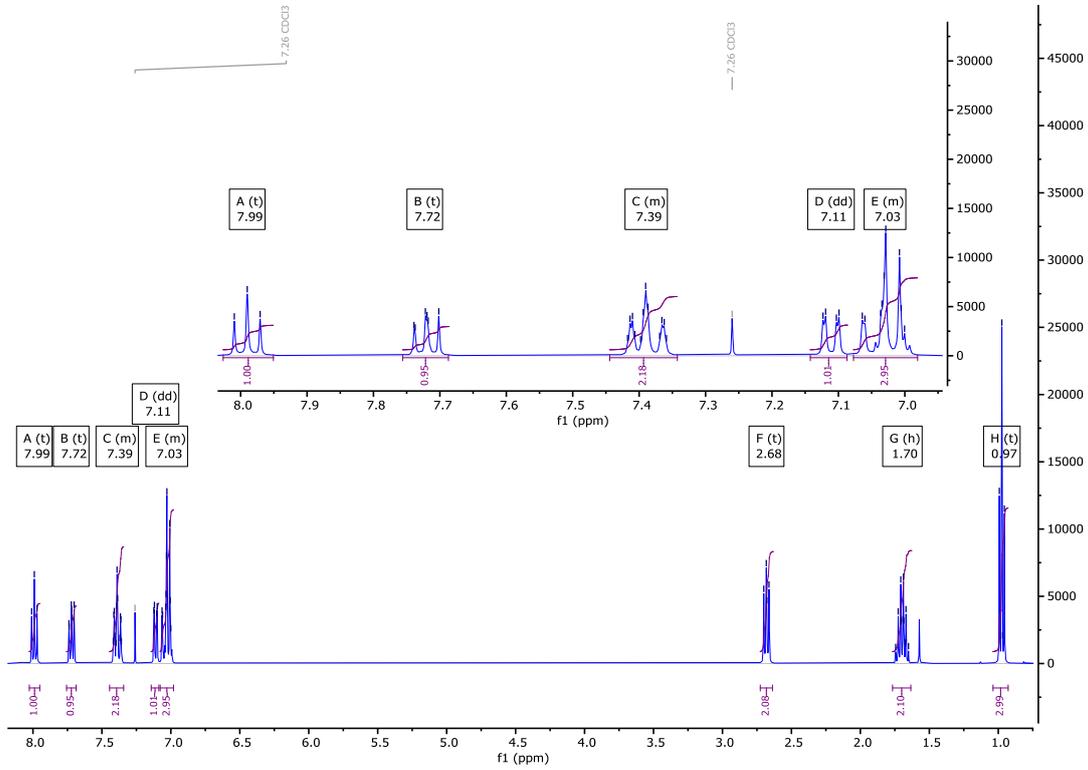

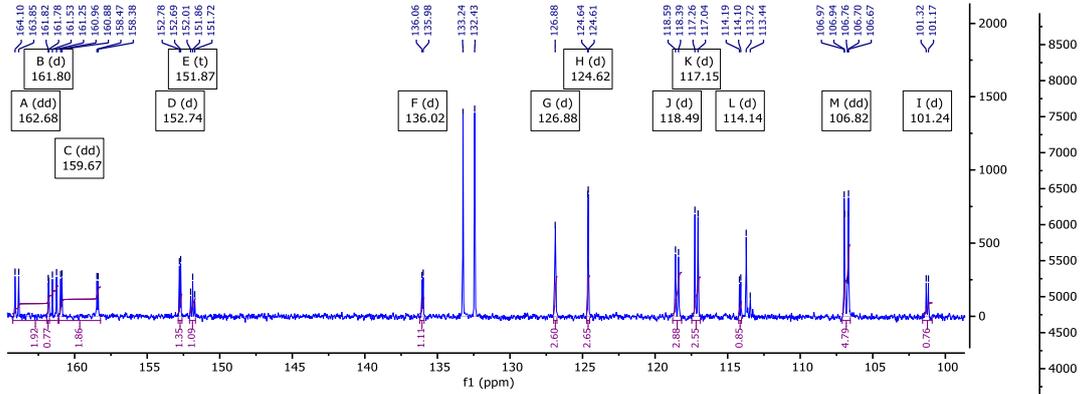

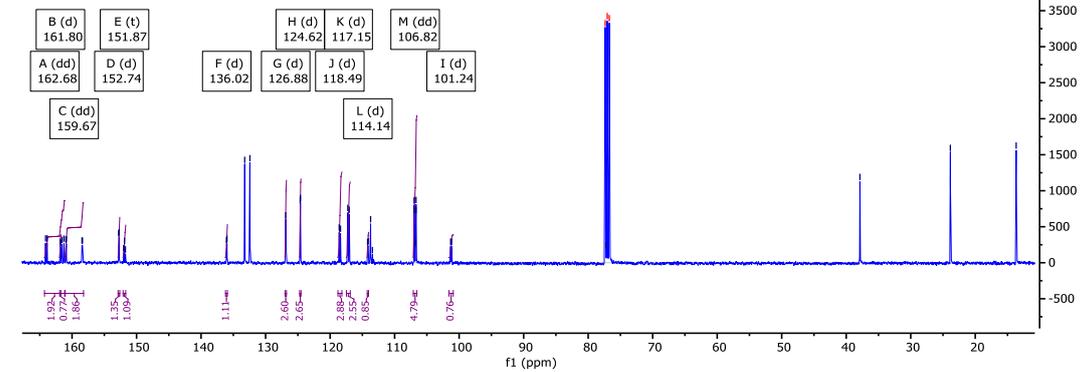

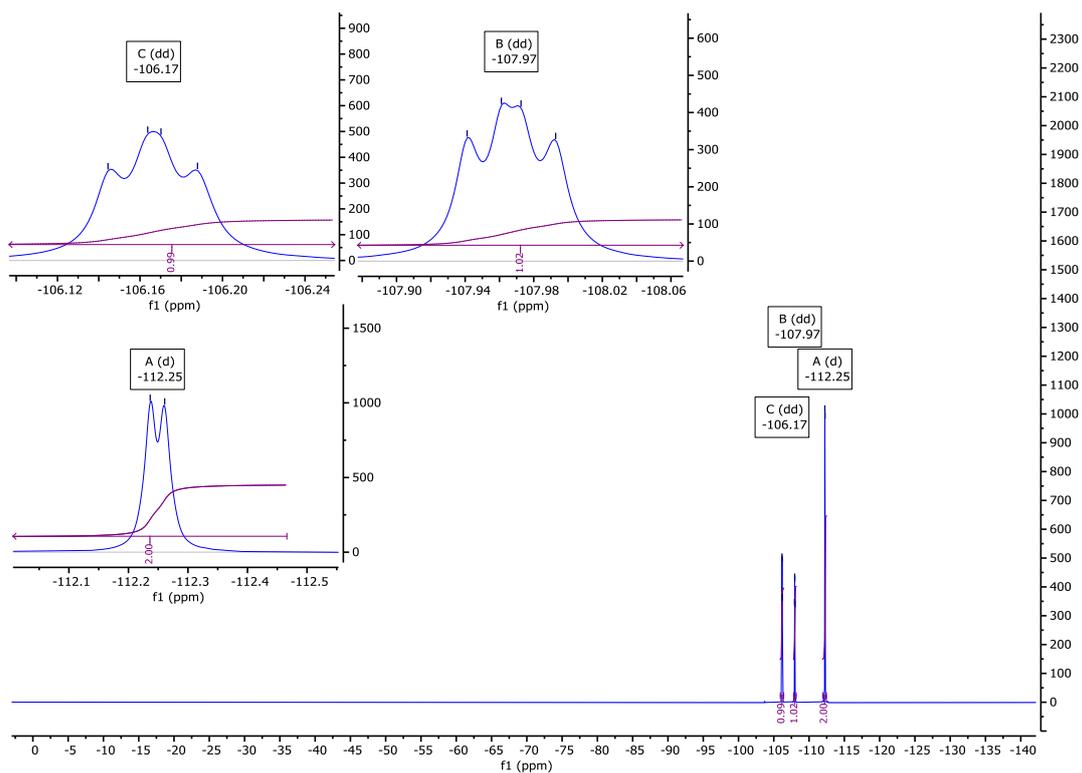

**Figure S8**:  Chemical structure, NMR spectra ($^1$H (top), $^{13}$C[1H] (middle), and $^{19}$F (bottom)) spectra for **8 (1.2.1)**.

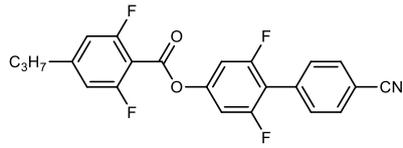

**11 (2.1.1)**

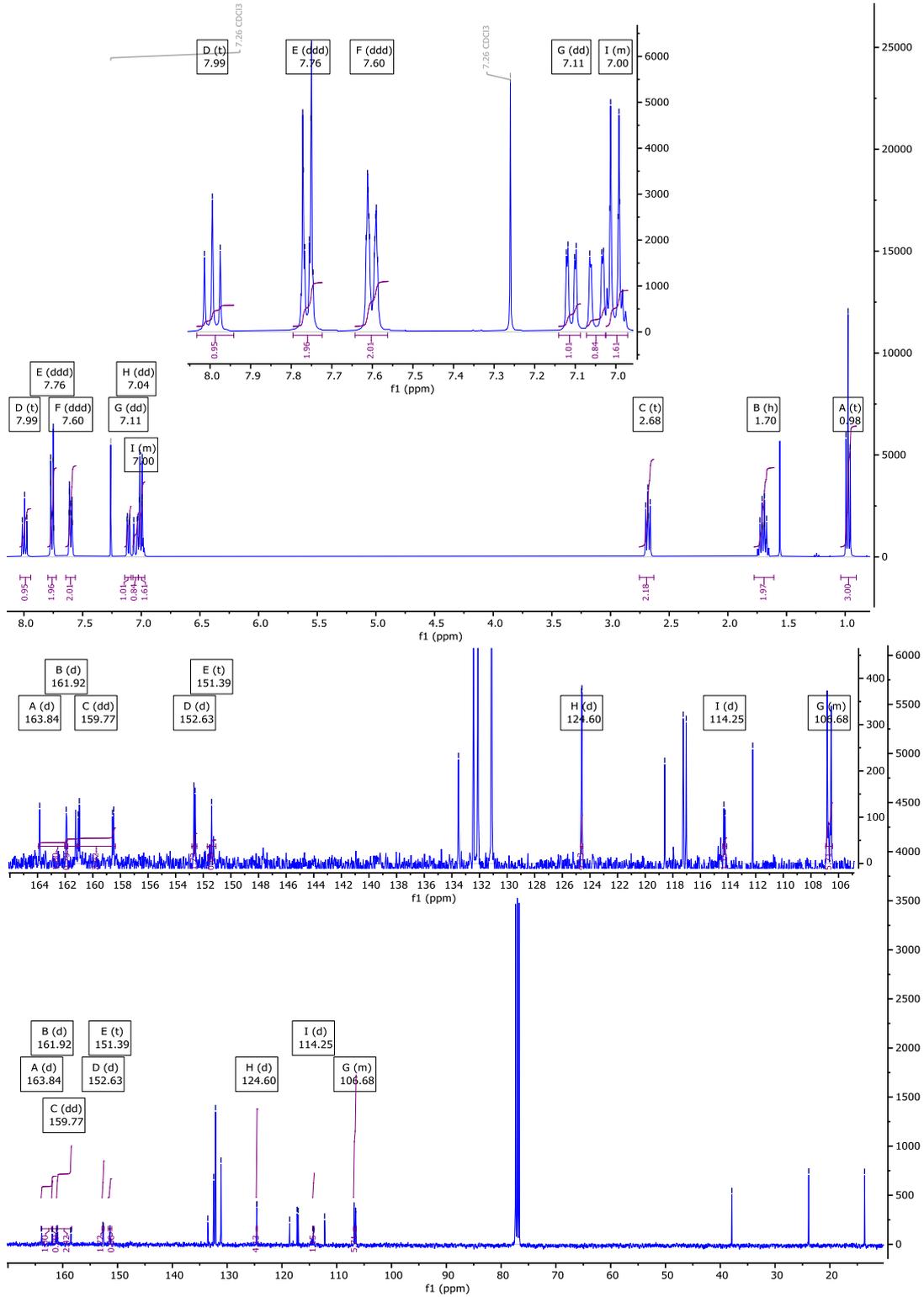

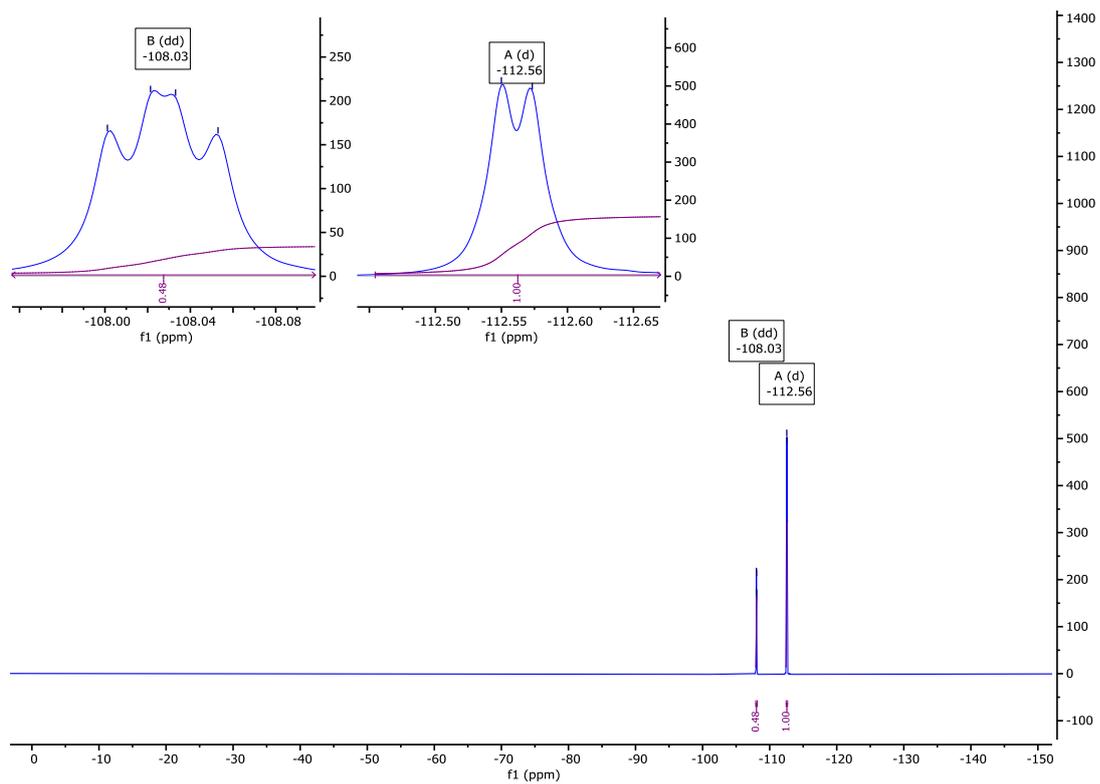

**Figure S9**: Chemical structure, NMR spectra (¹H (top), ¹³C[1H] (middle), and ¹⁹F (bottom)) spectra for **11 (0.2.1)**.

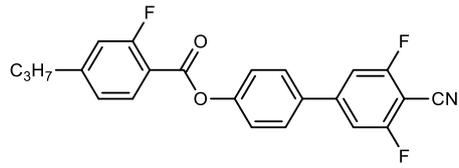

**17 (2.0.1)**

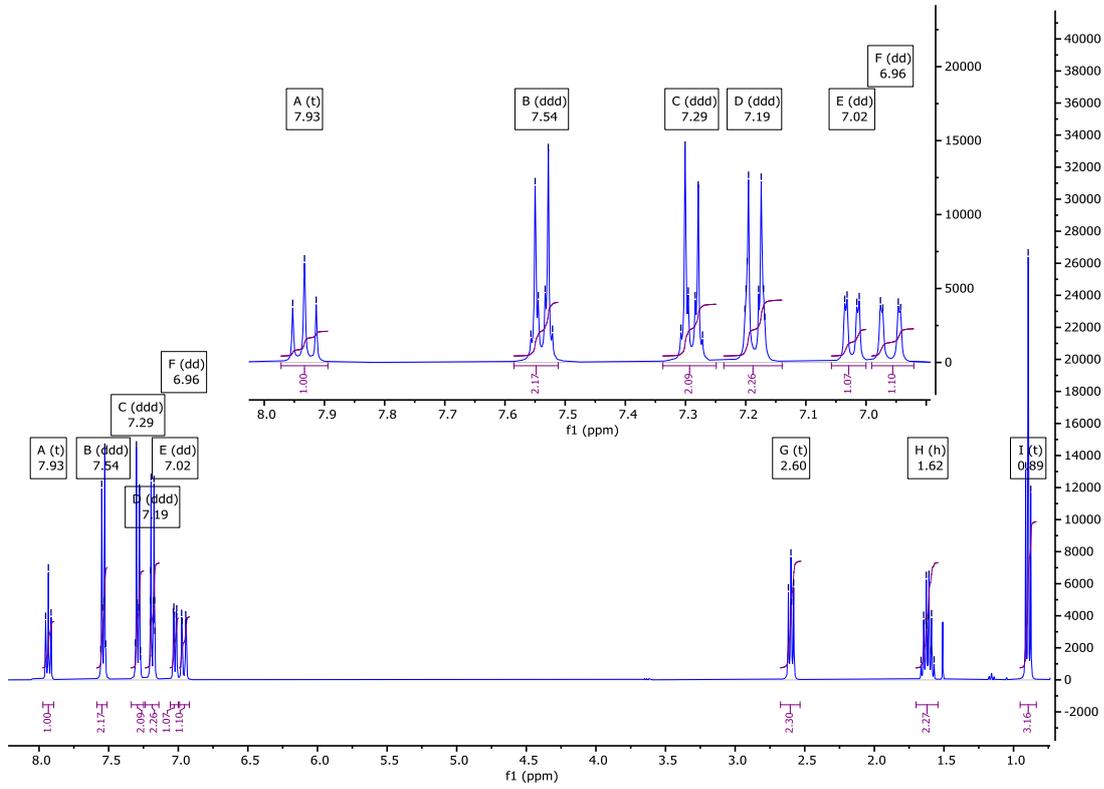

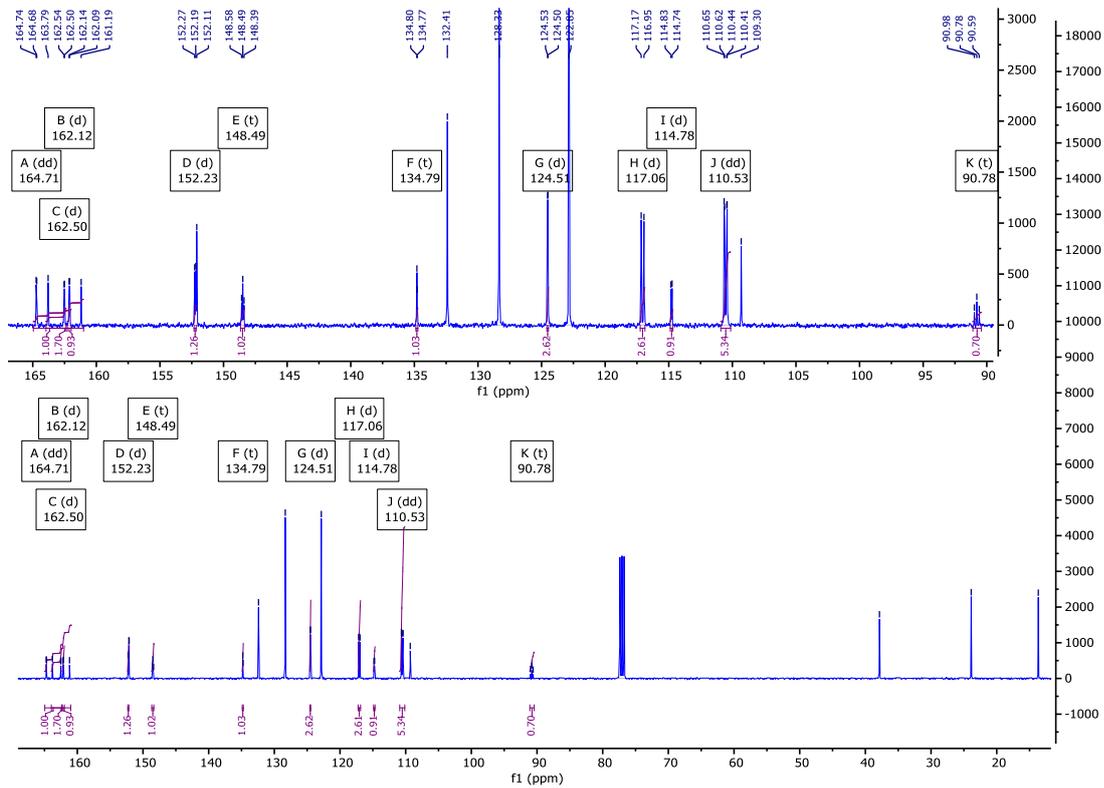

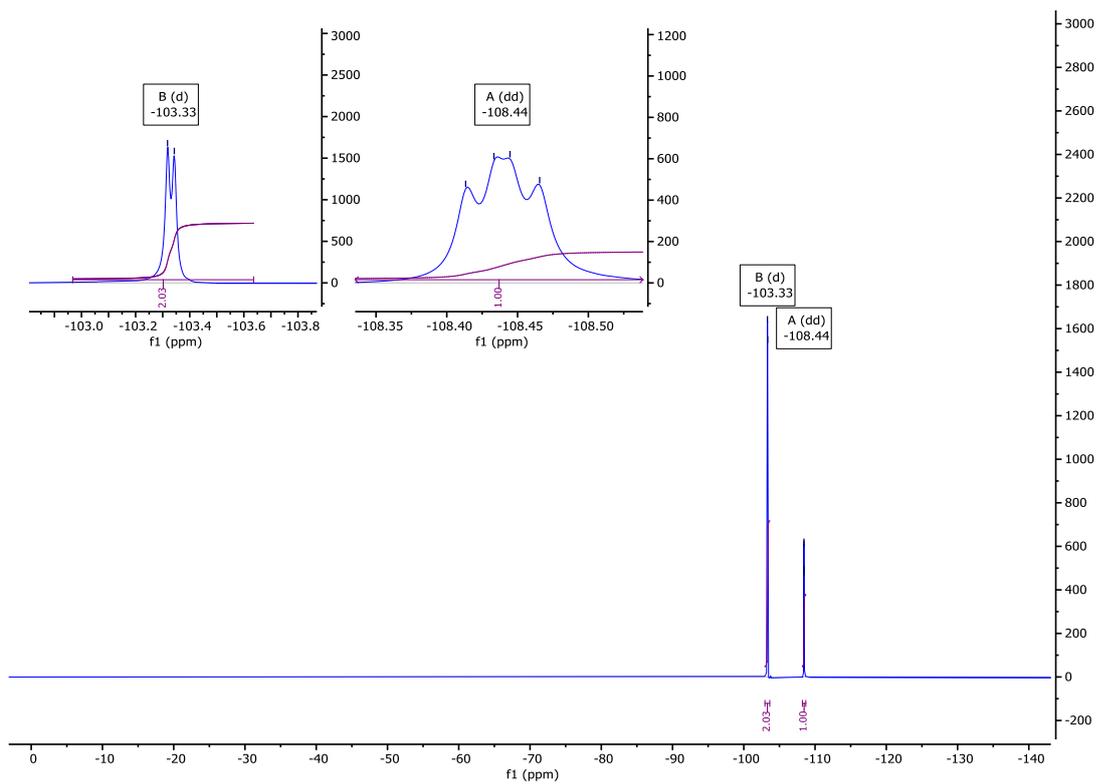

**Figure S10**: Chemical structure, NMR spectra ($^1$H (top), $^{13}$C[1H] (middle), and $^{19}$F (bottom)) spectra for **17 (2.0.1)**.

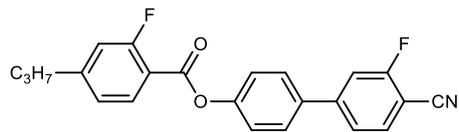

**20** (1.0.1)

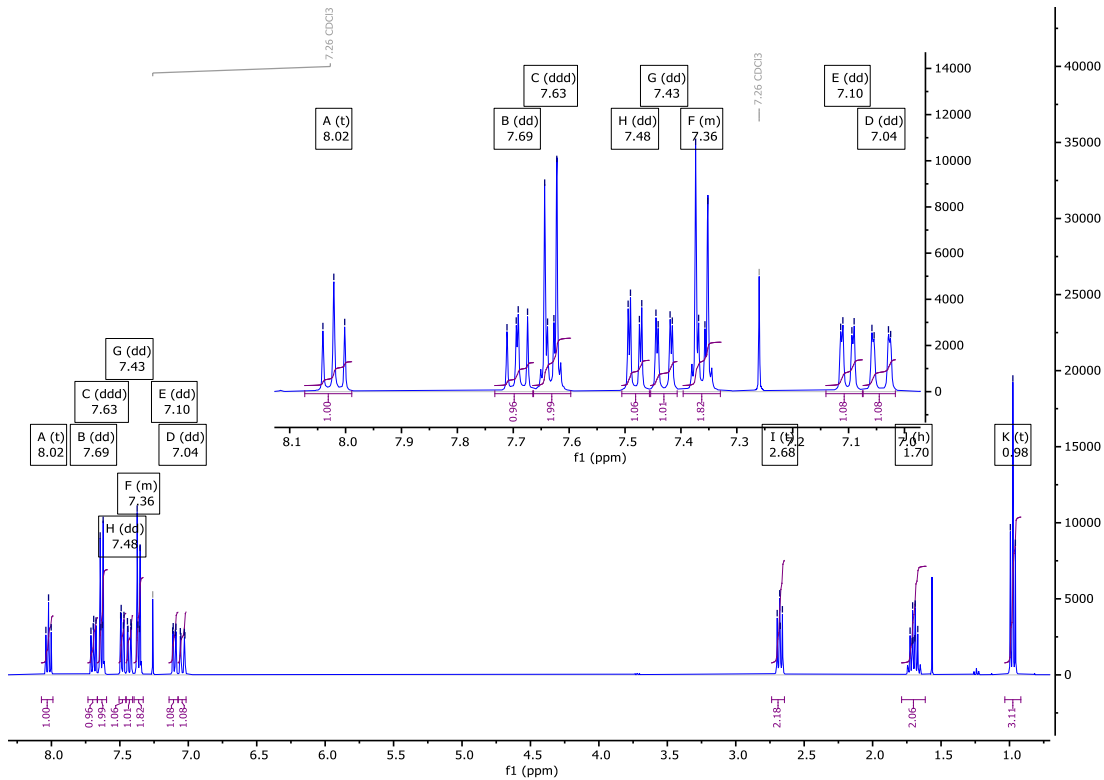

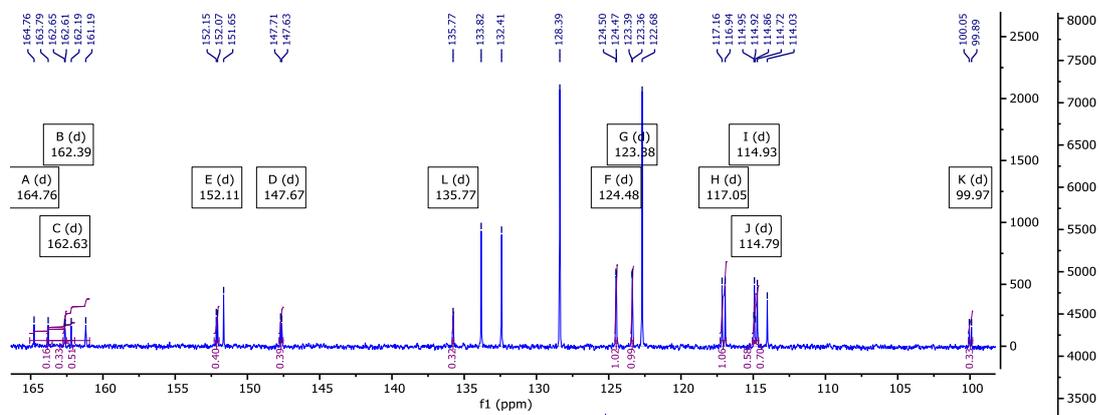

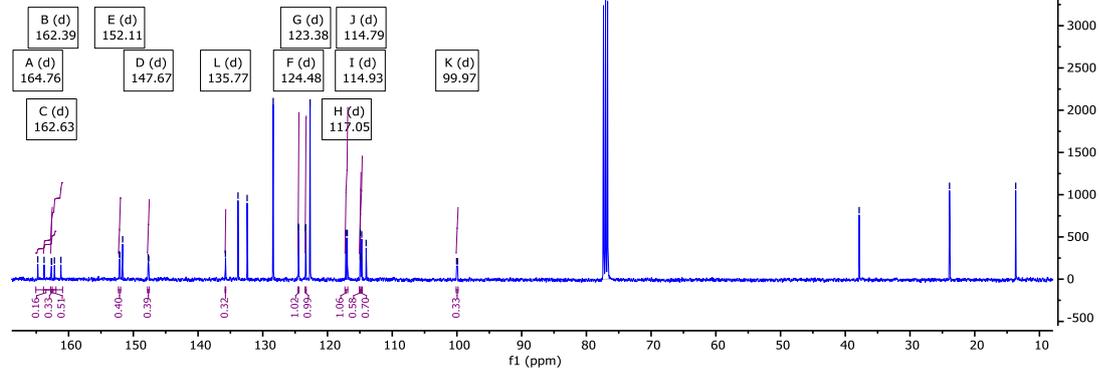

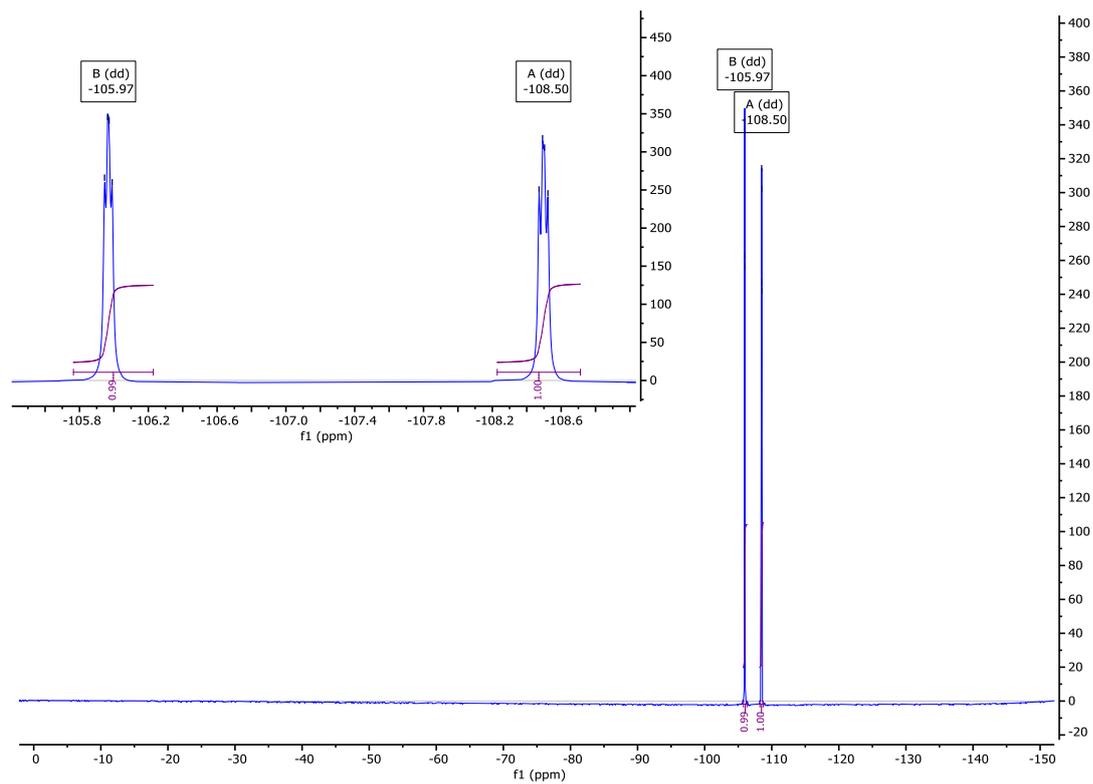

**Figure S11**: Chemical structure, NMR spectra ($^1$H (top), $^{13}$C[1H] (middle), and $^{19}$F (bottom)) spectra for **20 (1.0.1)**.

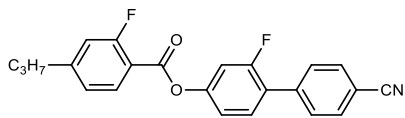

**23 (0.1.1)**

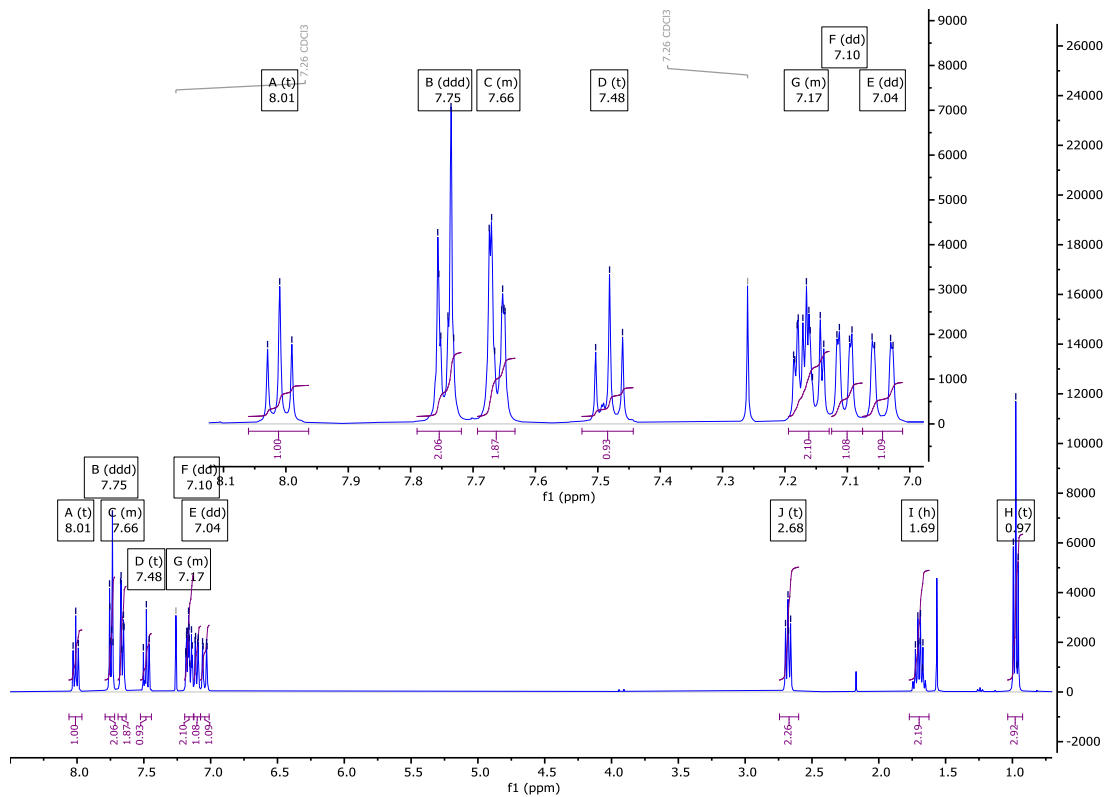

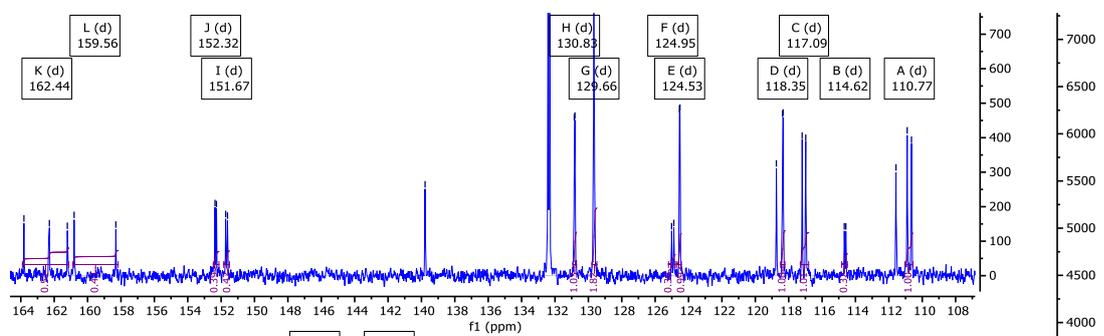

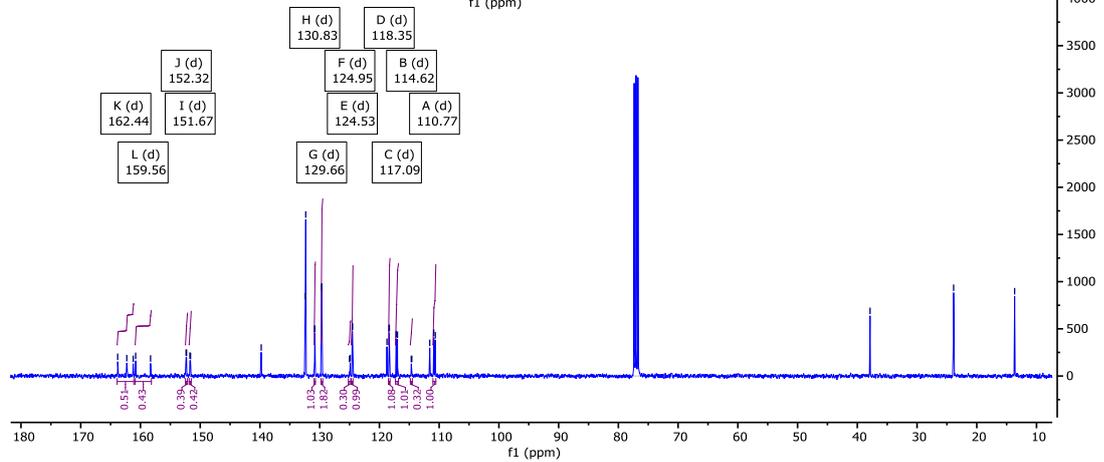

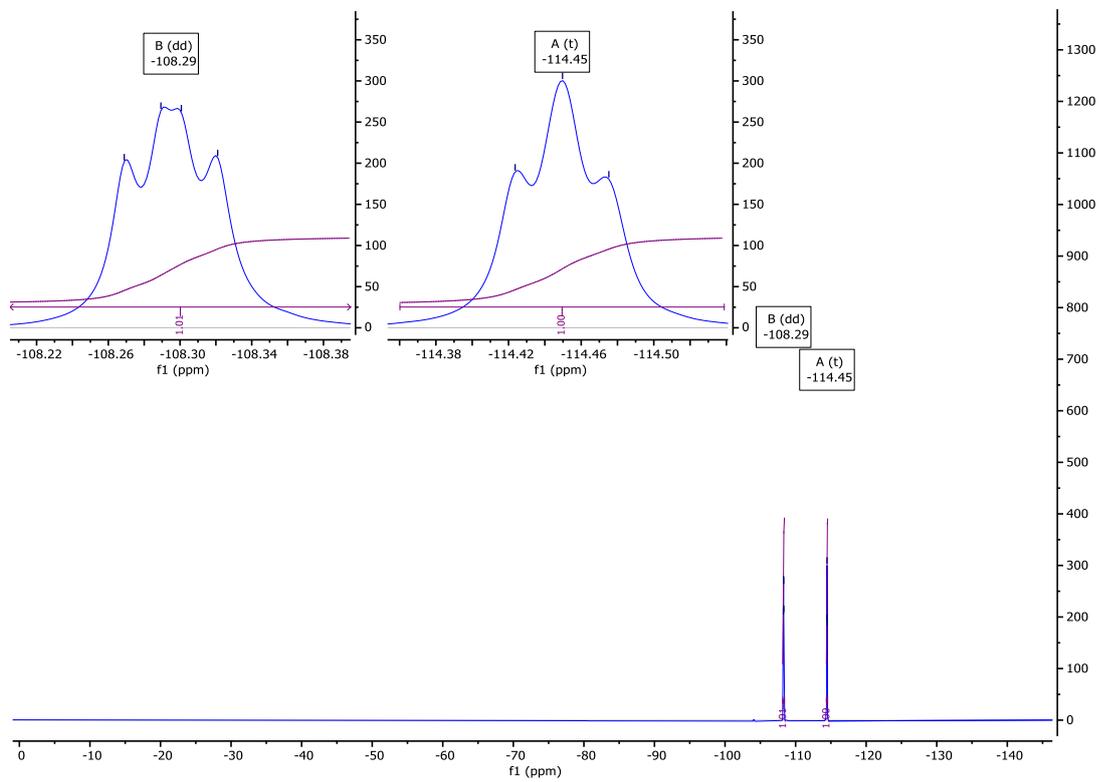

**Figure S12**: Chemical structure, NMR spectra ([1]H (top), [13]C[1H] (middle), and [19]F (bottom)) spectra for **23 (0.1.1)**.

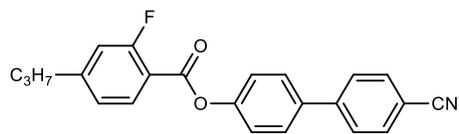

**26** (0.0.1)

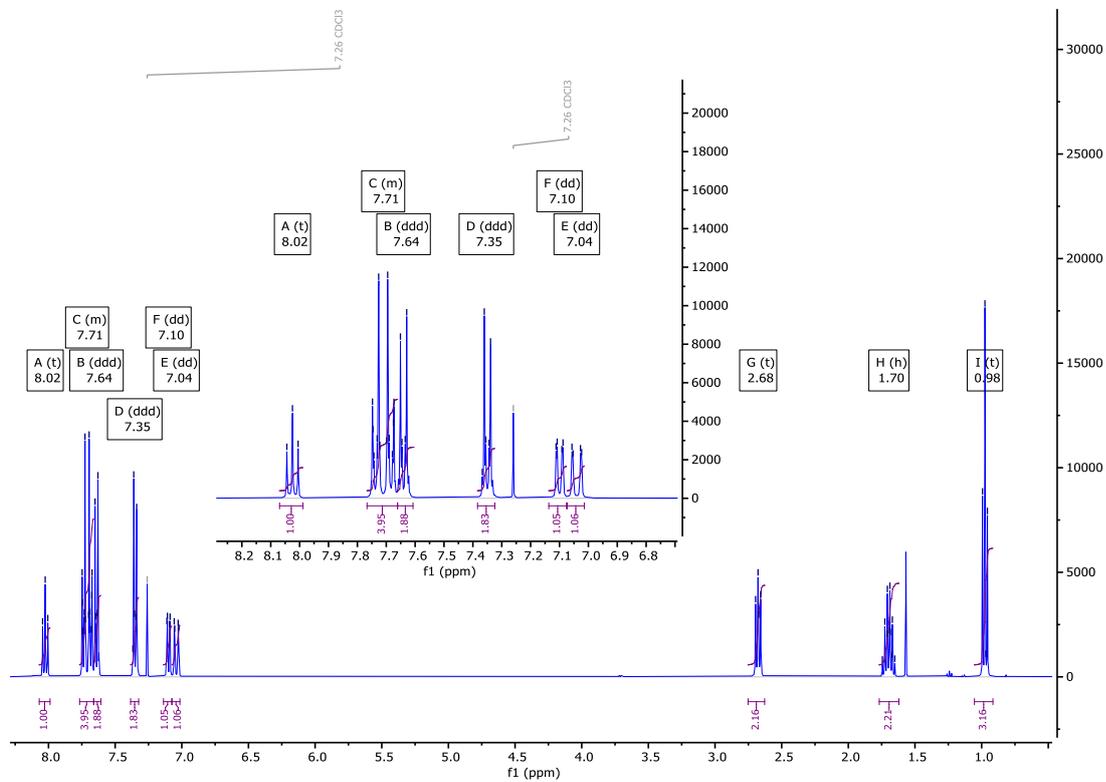

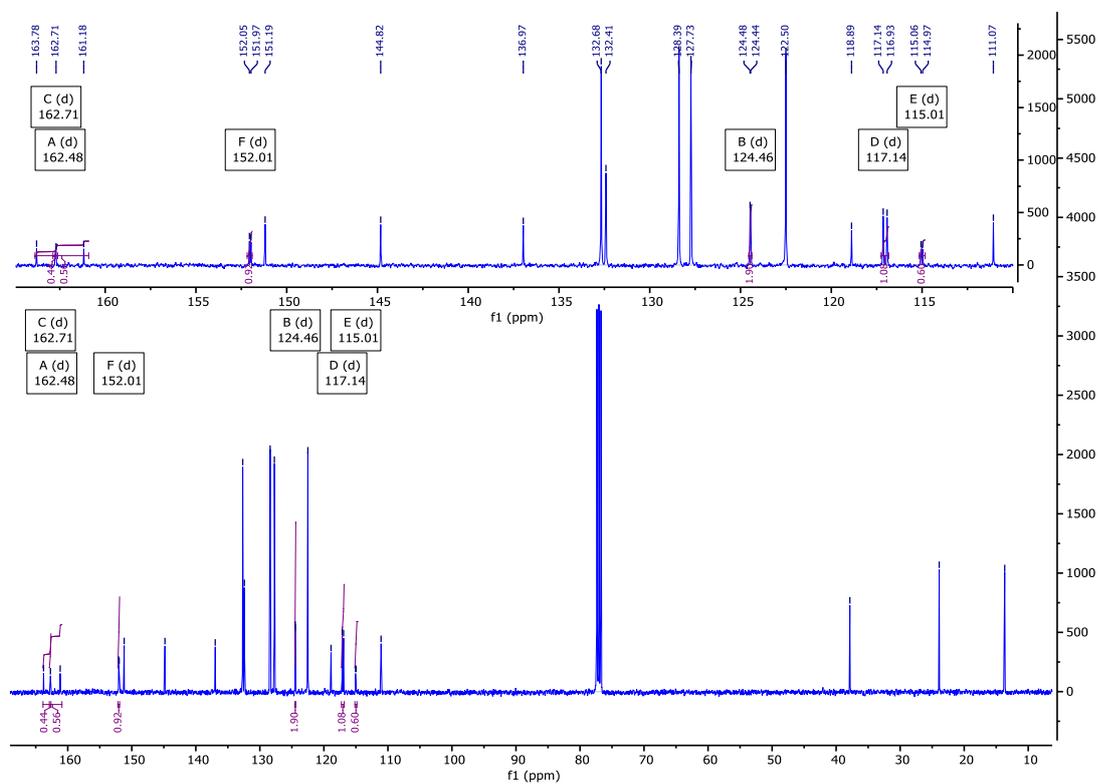

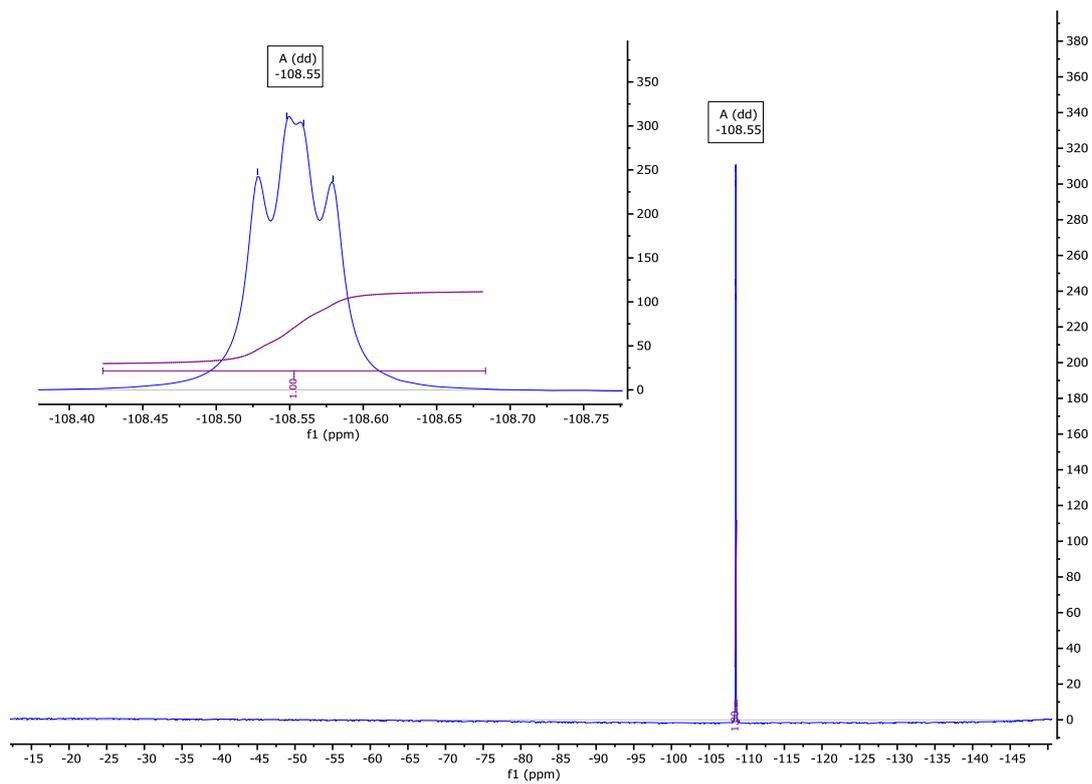

**Figure S13**: Chemical structure, NMR spectra ([1]H (top), [13]C[1H] (middle), and [19]F (bottom)) spectra for **26 (0.0.1)**.

## 4 Supplemental references